\documentclass[pra,twocolumn,amsmath,amssymb,amsfonts,longbibliography]{revtex4-1}
\usepackage{graphicx}
\usepackage{dcolumn}
\usepackage{bm}
\usepackage[colorlinks,urlcolor=blue,citecolor=blue,linkcolor=black]{hyperref}
\usepackage{amsmath}
\usepackage{amsfonts}
\usepackage{amssymb}
\usepackage{feynmp}
\usepackage{tabularx}
\usepackage{extarrows}%
\allowdisplaybreaks[1]

\begin{document}
\title{Consistent bosonization-debosonization II:\\\emph{The two-lead Kondo problem and the fate of its non-equilibrium Toulouse
point}}
\author{C.~J.~Bolech}
\affiliation{Department of Physics, University of Cincinnati, Ohio 45221-0011, USA}
\author{Nayana Shah}
\affiliation{Department of Physics, University of Cincinnati, Ohio 45221-0011, USA}
\date{August 12$^\mathrm{th}$, 2015}

\begin{abstract}
Following the development of a scheme to bosonize and debosonize consistently
[N. Shah and C.J. Bolech, Phys.~Rev B \textbf{93}, 085440 (2016); arXiv:1508.03078], we present in detail the
Toulouse-point analytic solution of the two-lead Kondo junction model. The
existence and location of the solvable point is not modified, but the
calculational methodology and the final expressions for observable quantities
change markedly as compared to the existent results. This solvable point is
one of the remarkably few exact results for nonequilibrium transport in
correlated systems. It yields relatively simple analytical expressions for the
current in the full range of temperature, magnetic field, and voltage. It also
shows precisely, within the limitations of the Toulouse fine-tuning, how the
transport evolves depending on the relative strengths of interlead and
intralead Kondo exchange couplings ranging from weak to strong. Thus its
improved understanding is an important stepping stone for future research.

\end{abstract}
\maketitle

\section{Introduction}

In the first part of this series \cite{shah2016}, we introduced a consistent
prescription in order to be able to bosonize, make transformations, and
debosonize consistently in the presence of ``active local impurities or
boundaries'', which we called the consistent bosonization-debosonization (BdB)
program. In this paper we explore the implications of this formalism for the
important case of quantum impurity problems.

Just over fifty years ago, in 1964, Kondo showed that a mysterious
finite-temperature minimum in the resistivity of metals was due to the
contributions from dilute magnetic impurities present in the samples
\cite{kondo1964}. This marked the start of the study of the so-called
\textit{Kondo problem}, which is one of the pillars of modern condensed-matter
theory \cite{JPSJ2005}. On the technical side, the reason for this is that the
problem of a single magnetic impurity in a metal is one of the very first
examples of an asymptotically free theory \cite{Cox&Z,*cox1998}. When the
system is below a certain energy scale known as the Kondo temperature, standard
perturbation theory fails and one needs to resort to more sophisticated
theoretical tools (many of which were actually first developed studying this
problem \cite{Hewson}). These range from the exact to the versatile, or from
the Bethe ansatz
\cite{andrei1980,wiegmann1980,kawakami1981,andrei1984,bolech2002,*bolech2005a}
to auxiliary-particle perturbation methods
\cite{abrikosov1965,barnes1976,*barnes1977,coleman1984,hafermann2009,munioz2013}. 
On the application side, the
relevance of the Kondo problem extends nowadays well beyond the original
system of impurities in metals. The so-called Kondo lattice is the central
model in the study of heavy fermions \cite{Cox&Z,*cox1998} and, even more
generally, a formalism know as dynamical mean field theory is based on the
mapping of any complicated tight-binding model of a material to an effective
quantum impurity problem \cite{georges1996}. Moreover, since the last two
decades, as the study of artificial mesoscopic systems reached the nanoscale,
the Kondo problem can show up in all sorts of electronic devices, most notably
in semiconductor quantum dots \cite{goldhaber1998,cronenwett1998} but also in
molecular electronics \cite{yu2004}, etc.

The experiments with artificial nanostructures brought in an additional layer
of complication to the theory of the Kondo effect. Most of the typical
experiments involve transport measurements in nonequilibrium conditions,
while our best theoretical tools to describe systems out of equilibrium are
perturbative. Since the Kondo effect is nonperturbative, a lot of theoretical
activity continues to ensue, to the point that it is fair to say that a deep
understanding of correlated systems out of equilibrium is still work in progress.

Early on during the nanoscale revolution in mesoscopics, the pioneering
theoretical work of Schiller and Hershfield (S\&H) provided the first (and
still now one of the few) exact solution of a nonequilibrium strongly
correlated quantum problem \cite{schiller1995}. A few years before them, Emery
and Kivelson (E\&K) found a solvable point (called a Toulouse point) for the
two-channel Kondo model \cite{emery1992} (in which, besides spin, the band
electrons have one more two-valued discrete degree of freedom
\cite{nozieres1980}). S\&H realized that they could adapt it to the case of a
Kondo impurity interacting with two separate leads. They thus found a mapping
that allows for the calculation of transport in a problem that has strong
correlations due to the exchange interaction between the electrons and the
impurity spin. We will revisit the central aspects of their work below. We
shall find that while the key insights of S\&H about the existence of a
nonequilibrium Toulouse point remain valid, our generic
consistent-debosonization procedure shows that the actual observables being
calculated are substantially modified and yield more physically consistent results.

\section{Model of a Kondo Junction}

We shall be interested in a class of systems in which the transport between
two leads or terminals happens across a microscopic region in which the
fermionic degrees of freedom are such that the region has a total magnetic
moment that remains unscreened (and we will focus on the case of spin-1/2).
This situation is typical of nano-scale quantum dots with strong Coulomb
blockade and has been an experimental reality since the late 1990s
\cite{goldhaber1998,cronenwett1998}. At low temperatures, remarkably, the
system enters the so-called Kondo regime and is able to conduct despite the
Coulomb blockade.

\begin{figure}[ptb]
\begin{center}
\includegraphics[width=\columnwidth]{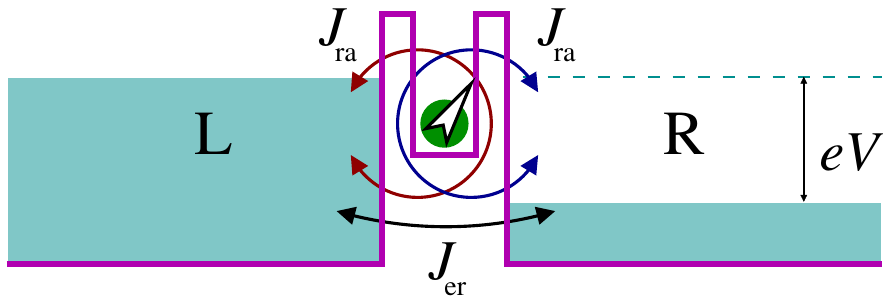}
\end{center}
\caption{Schematic depiction of the setting in which two Fermi seas kept at
different chemical potentials with their difference given by $eV=\mu
_{\text{L}}-\mu_{\text{R}}$ are connected via quantum tunneling across the
potential barrier that separates them. The situation when the barrier region
does not allow for internal states was discussed before \cite{shah2016}. Here
there exists a many-body state trapped by the barrier with an unscreened total
spin $1/2$ that we refer to as the impurity (depicted by an arrowhead). Due to
a strong Coulomb blockade, the impurity interacts with the electrons in the
leads via exchange processes only. For the purpose of the figure we
generically denote by $J_{\text{ra}}$ the int\underline{ra}-lead exchange
terms (either parallel or perpendicular) and by $J_{\text{er}}$ the
int\underline{er}-lead co\-tunneling exchange processes that can give rise to a
current (a similar notation will be introduced in the text later). }%
\label{Fig_dot}%
\end{figure}

We are thus interested in the low-temperature characteristics of nonlinear
transport across a quantum dot in the Kondo regime. We will model the system
with a two-lead version of the Kondo model (see Fig.~\ref{Fig_dot}). 
In analogy to the equilibrium case, 
this model can be derived from a more microscopic Anderson-type model
via a (time-dependent) Schrieffer-Wolff transformation \cite{Hewson,
kaminski2000}. In Hamiltonian language, the model is given by%

\begin{subequations}
\label{Eq:2LK}%
\begin{align}
H  &  =\sum_{\sigma,\ell}\left(  \int\mathcal{H}_{\ell}^{0}\,dx+H_{\text{K}%
}^{z}+H_{\text{K}}^{\perp}\right)  +H_{\text{field}}\,\text{,}\\
\mathcal{H}_{\ell}^{0}  &  =\psi_{\sigma\ell}^{\dagger}\left(  x,t\right)
\left(  -iv_{\text{F}}\partial_{x}\right)  \psi_{\sigma\ell}\left(
x,t\right)  \,\text{,}\\
H_{\text{K}}^{z}  &  =J_{\ell\ell^{\prime}}^{z}\,S_{\text{imp}}^{z}\left(
\frac{\sigma}{2}\psi_{\sigma\ell}^{\dagger}\left(  0,t\right)  \psi
_{\sigma\ell^{\prime}}\left(  0,t\right)  \right)  \,\text{,}\\
H_{\text{K}}^{\perp}  &  =J_{\ell\ell^{\prime}}^{\perp}\,S_{\text{imp}%
}^{\sigma}\left(  \frac{1}{2}\psi_{\bar{\sigma}\ell}^{\dagger}\left(
0,t\right)  \psi_{\sigma\ell^{\prime}}\left(  0,t\right)  \right)
\,\text{,}\\
H_{\text{field}}  &  =-h\,S_{\text{imp}}^{z}\,\text{,}%
\end{align}
where $\psi_{\sigma\ell}\left(  x,t\right)  $ are chiral fermions in the
Heisenberg representation that are obtained after unfolding the two leads in
the usual way \cite{affleck1995,shah2016}. We adopt the notation
$\sigma=\left\{  \downarrow,\uparrow\right\}  =\left\{  -1,+1\right\}  $ and
$\ell=\left\{  \text{L},\text{R}\right\}  =\left\{  -1,+1\right\}$ for the
spin and lead index, respectively. The bar notation denotes a sign 
change---\textrm{e.g.}, $\bar{\sigma}=-\sigma$. The impurity is described by
$S_{\text{imp}}^{z}$ and $S_{\text{imp}}^{\sigma}=S_{\text{imp}}^{x}+i\sigma
S_{\text{imp}}^{y}$. We assume now that at a much earlier time the connection
between the two leads was established and that there is a battery keeping a
constant chemical-potential difference between the two leads (what we call a
Landauer-type configuration \cite{imry1999,blanter2000}). Under these
conditions the system will be in a nonequilibrium steady state
\cite{doyon2006}. Let us call $\mu_{\ell}$ the chemical potential of lead
$\ell$, such that $\mu_{\text{L}}-\mu_{\text{R}}=eV$, with $V$ the voltage
drop across the junction. The information about these chemical potentials will
enter into the distribution functions for each lead.

\section{Bosonization-Debosonization Approach}

To set the stage for the bosonization of the model, we start by gauging away
the chemical-potential difference. For that, let us first switch to Lagrangian
language in which the system is described by%
\end{subequations}
\begin{subequations}
\begin{align}
\mathcal{L}_{\ell}^{0}  &  =\psi_{\sigma\ell}^{\dagger}\left(  x,t\right)
\left(  i\partial_{t}\right)  \psi_{\sigma\ell}\left(  x,t\right)
-\mathcal{H}_{\ell}^{0}\nonumber\\
&  =\psi_{\sigma\ell}^{\dagger}\left(  x,t\right)  \left(  i\partial
_{t}+iv_{\text{F}}\partial_{x}\right)  \psi_{\sigma\ell}\left(  x,t\right)
\,\text{,}\\
L_{\text{K}}^{z}  &  =-H_{\text{K}}^{z}=-J_{\ell\ell^{\prime}}^{z}%
\,S_{\text{imp}}^{z}\left(  \frac{\sigma}{2}\psi_{\sigma\ell}^{\dagger}\left(
0,t\right)  \psi_{\sigma\ell^{\prime}}\left(  0,t\right)  \right)
\,\text{,}\\
L_{\text{K}}^{\perp}  &  =-H_{\text{K}}^{\perp}=-J_{\ell\ell^{\prime}}^{\perp
}\,S_{\text{imp}}^{\sigma}\left(  \frac{1}{2}\psi_{\bar{\sigma}\ell}^{\dagger
}\left(  0,t\right)  \psi_{\sigma\ell^{\prime}}\left(  0,t\right)  \right)
\,\text{.}%
\end{align}
We can now make the following (gauge) field transformation $\psi_{\sigma\ell
}\left(  x,t\right)  =e^{-i\mu_{\ell}t}\check{\psi}_{\sigma\ell}\left(
x,t\right)  $. The important point is that now the distribution functions do
not contain information about the chemical potentials any longer (cf.~with the
discussion for the case of a simple junction \cite{shah2016}). Next we
subtract the vev (vacuum expectation value), which for a noninteracting
problem is equivalent to factoring out the fast oscillations in each lead
according to $\check{\psi}_{\sigma\ell}\left(  x,t\right)  =e^{ik_{\text{F}%
}^{\ell}x}\breve{\psi}_{\sigma\ell}\left(  x,t\right)  $, with $k_{\text{F}%
}^{\ell}=\mu_{\ell}/v_{\text{F}}$ for this linear-dispersion case. So we are
naturally, thanks to the linear dispersion, lead to the normal-ordered
formulation of the problem:%
\end{subequations}
\begin{subequations}
\begin{align}
\mathcal{L}_{\ell}^{0}  &  =\,\colon\breve{\psi}_{\sigma\ell}^{\dagger}\left(
x,t\right)  \left(  i\partial_{t}+iv_{\text{F}}\partial_{x}\right)
\breve{\psi}_{\sigma\ell}\left(  x,t\right)  \colon\,\text{,}\\
L_{\text{K}}^{z}  &  =-J_{\ell\ell}^{z}\,S_{\text{imp}}^{z}\left(
\frac{\sigma}{2}\colon\breve{\psi}_{\sigma\ell}^{\dagger}\left(  0,t\right)
\breve{\psi}_{\sigma\ell}\left(  0,t\right)  \colon\right)  -\nonumber\\
&  \quad-e^{i\bar{\ell}eVt}J_{\ell\bar{\ell}}^{z}\,S_{\text{imp}}^{z}\left(
\frac{\sigma}{2}\breve{\psi}_{\sigma\ell}^{\dagger}\left(  0,t\right)
\breve{\psi}_{\sigma\bar{\ell}}\left(  0,t\right)  \right)  \,\text{,}\\
L_{\text{K}}^{\perp}  &  =-J_{\ell\ell}^{\perp}\,S_{\text{imp}}^{\sigma
}\left(  \frac{1}{2}\breve{\psi}_{\bar{\sigma}\ell}^{\dagger}\left(
0,t\right)  \breve{\psi}_{\sigma\ell}\left(  0,t\right)  \right)  -\nonumber\\
&  \quad-e^{i\bar{\ell}eVt}J_{\ell\bar{\ell}}^{\perp}\,S_{\text{imp}}^{\sigma
}\left(  \frac{1}{2}\breve{\psi}_{\bar{\sigma}\ell}^{\dagger}\left(
0,t\right)  \breve{\psi}_{\sigma\bar{\ell}}\left(  0,t\right)  \right)
\,\text{.}%
\end{align}
Notice that in the case of the \textit{parallel} intralead impurity terms we
wrote them also as normal ordered (since the vev's of the two spin projections
cancel each other due to the $\sigma/2$ factor), which is customary in other
approaches such as boundary conformal field theory (BCFT), whereas the inter\-lead
impurity terms naturally remain non-normal-ordered. At this point we lost the
information about the absolute energy reference but we still have the
information about the potential drop encoded in the time-dependent phase of
the tunneling term.

\subsection{Bosonization and Initial Mappings}

The first part of the BdB program starts by bosonizing, of course. In order to
do that, we go back to the Hamiltonian formulation of the problem,%
\end{subequations}
\begin{subequations}
\begin{align}
\mathcal{H}_{\ell}^{0}  &  =\,\colon\breve{\psi}_{\sigma\ell}^{\dagger}\left(
x,t\right)  \left(  -iv_{\text{F}}\partial_{x}\right)  \breve{\psi}%
_{\sigma\ell}\left(  x,t\right)  \colon\,\text{,}\\
H_{\text{K}}^{z}  &  =J_{\ell\ell}^{z}\,S_{\text{imp}}^{z}\left(  \frac
{\sigma}{2}\colon\breve{\psi}_{\sigma\ell}^{\dagger}\left(  0,t\right)
\breve{\psi}_{\sigma\ell}\left(  0,t\right)  \colon\right)  +\nonumber\\
&  \quad+e^{i\bar{\ell}eVt}J_{\ell\bar{\ell}}^{z}\,S_{\text{imp}}^{z}\left(
\frac{\sigma}{2}\breve{\psi}_{\sigma\ell}^{\dagger}\left(  0,t\right)
\breve{\psi}_{\sigma\bar{\ell}}\left(  0,t\right)  \right)  \,\text{,}\\
H_{\text{K}}^{\perp}  &  =J_{\ell\ell}^{\perp}\,S_{\text{imp}}^{\sigma}\left(
\frac{1}{2}\breve{\psi}_{\bar{\sigma}\ell}^{\dagger}\left(  0,t\right)
\breve{\psi}_{\sigma\ell}\left(  0,t\right)  \right)  +\nonumber\\
&  \quad+e^{i\bar{\ell}eVt}J_{\ell\bar{\ell}}^{\perp}\,S_{\text{imp}}^{\sigma
}\left(  \frac{1}{2}\breve{\psi}_{\bar{\sigma}\ell}^{\dagger}\left(
0,t\right)  \breve{\psi}_{\sigma\bar{\ell}}\left(  0,t\right)  \right)
\,\text{;}%
\end{align}
and we bosonize according to $\mathcal{H}_{\ell}^{0}$, with $H_{\text{K}%
}^{z,\perp}$ taken as the interaction terms. We follow the same standard
bosonization prescription as we did for the junction problem \cite{shah2016},
$\breve{\psi}_{\sigma\ell}\left(  x,t\right)  =\frac{1}{\sqrt{2\pi a}%
}F_{\sigma\ell}\left(  t\right)  e^{-i\phi_{\sigma\ell}\left(  x,t\right)  }$,
and in terms of the bosons the Hamiltonian density for the leads takes the
usual form (the Klein factors, $F_{\sigma\ell}$, drop out from these terms)%
\end{subequations}
\begin{equation}
\mathcal{H}^{0}=\sum_{\ell}\mathcal{H}_{\ell}^{0}=\frac{v_{\text{F}}}{4\pi
}\sum_{\substack{\sigma=\uparrow,\downarrow\\\ell=\text{L,R}}}\colon\left[
\partial_{x}\phi_{\sigma\ell}\left(  x,t\right)  \right]  ^{2}\colon\,\text{.}%
\end{equation}
Using the same standard physically-motivated rotated boson basis,
$\phi_{\sigma\ell}=\left(  \phi_{c}+\sigma\phi_{s}+\ell\phi_{l}+\sigma\ell
\phi_{sl}\right)  /2$, as we did for the simple junction \cite{shah2016}, the
non-interacting Hamiltonian density retains its quadratic form%
\begin{equation}
\mathcal{H}^{0}=\frac{v_{\text{F}}}{4\pi}\sum_{\nu=c,s,l,sl}\colon\left[
\partial_{x}\phi_{\nu}\left(  x,t\right)  \right]  ^{2}\colon\,\text{.}%
\end{equation}

Let us postpone the discussion of $H_{\text{K}}^{z}$, and proceed to bosonize
the ``perpendicular'' part of the Kondo term. The first part is the intralead
one, or lead-nonmixing, and it is present also in the standard two-channel
Kondo model, while the second part is interlead, or lead-mixing, and is
responsible for transport as can already be seen from the voltage dependence
(we keep the time and space dependence of the bosonic fields implicit for the
sake of brevity):%
\begin{align}
H_{\text{K}}^{\perp}  &  =\frac{J_{\ell\ell}^{\perp}}{2\pi a}\,S_{\text{imp}%
}^{\sigma}\left(  \frac{1}{2}F_{\bar{\sigma}\ell}^{\dagger}e^{i\phi
_{\bar{\sigma}\ell}}F_{\sigma\ell}e^{-i\phi_{\sigma\ell}}\right)  +\nonumber\\
&  \quad+e^{i\bar{\ell}eVt}\frac{J_{\ell\bar{\ell}}^{\perp}}{2\pi
a}\,S_{\text{imp}}^{\sigma}\left(  \frac{1}{2}F_{\bar{\sigma}\ell}^{\dagger
}e^{i\phi_{\bar{\sigma}\ell}}F_{\sigma\bar{\ell}}e^{-i\phi_{\sigma\bar{\ell}}%
}\right)  \,\text{.}%
\end{align}
We now change to the rotated boson basis and being careful of not combining
vertex operators with opposite signs we introduce $\tilde{n}$ factors in the
same manner as in our consistent BdB treatment of the simple junction problem
\cite{shah2016}; notice as well that we are also introducing the $1/2$ factors
associated with consistent boundary conditions (CBCs) \cite{shah2016}. We get%
\begin{align}
H_{\text{K}}^{\perp}  &  =\frac{J_{\ell\ell}^{\perp}}{2\pi a}\frac{\tilde
{n}_{c}\tilde{n}_{l}^{\ell}}{2}\,S_{\text{imp}}^{\sigma}\left(  F_{\bar
{\sigma}\ell}^{\dagger}F_{\sigma\ell}e^{i\bar{\sigma}\phi_{s}}e^{i\bar{\sigma
}\ell\phi_{sl}}\right)  +\nonumber\\
&  \quad+e^{i\bar{\ell}eVt}\frac{J_{\ell\bar{\ell}}^{\perp}}{2\pi a}%
\frac{\tilde{n}_{c}\tilde{n}_{sl}^{\bar{\sigma}\ell}}{2}\,S_{\text{imp}%
}^{\sigma}\left(  F_{\bar{\sigma}\ell}^{\dagger}F_{\sigma\bar{\ell}}%
e^{i\bar{\sigma}\phi_{s}}e^{i\ell\phi_{l}}\right)  \,\text{,}%
\end{align}
where almost all possible $\tilde{n}$ factors, \textrm{i.e.},
\begin{subequations}
\label{Eq:Ntwiddles}%
\begin{align}
\tilde{n}_{c}  &  \equiv e^{i\phi_{c}/2}e^{-i\phi_{c}/2}/\sqrt{2}\,\text{,}\\
\tilde{n}_{s}^{\sigma}  &  \equiv e^{i\sigma\phi_{s}/2}e^{-i\sigma\phi_{s}%
/2}/\sqrt{2}\,\text{,}\\
\tilde{n}_{l}^{\ell}  &  \equiv e^{i\ell\phi_{l}/2}e^{-i\ell\phi_{l}/2}%
/\sqrt{2}\,\text{,}\\
\tilde{n}_{sl}^{\sigma\ell}  &  \equiv e^{i\sigma\ell\phi_{sl}/2}%
e^{-i\sigma\ell\phi_{sl}/2}/\sqrt{2}\,\text{,}%
\end{align}
appear except for the ones from the \textit{spin} sector.

The same as in the case of the simple barrier junction, we do not expect the
Klein factors to modify the physics. We treat them as we did in that case
\cite{vondelft1998,*zarand2000,bolech2006a,*iucci2008} by identifying
relations between different bilinears of \textit{original} and \textit{new}
Klein factors and fixing the four arbitrary phases; see Eqs.~(16a)-(16d) from
Ref.~%
\onlinecite{shah2016}%
. The rest of the Klein-factor relations can be derived from these
\cite{lee2002}. In particular, for the intra\-lead terms, we need
\end{subequations}
\begin{subequations}
\begin{align}
F_{\uparrow\text{R}}^{\dagger}F_{\downarrow\text{R}}  &  =F_{sl}^{\dagger
}F_{s}^{\dagger}\,\text{,}\label{Eq:KleinA}\\
F_{\uparrow\text{L}}^{\dagger}F_{\downarrow\text{L}}  &  =F_{sl}F_{s}%
^{\dagger}\,\text{,}\\
F_{\downarrow\text{R}}^{\dagger}F_{\uparrow\text{R}}  &  =F_{s}F_{sl}%
\,\text{,}\\
F_{\downarrow\text{L}}^{\dagger}F_{\uparrow\text{L}}  &  =F_{s}F_{sl}%
^{\dagger}\,\text{,}%
\end{align}
whereas for the inter\-lead terms we need%
\begin{align}
F_{\uparrow\text{R}}^{\dagger}F_{\downarrow\text{L}}  &  =F_{s}^{\dagger}%
F_{l}^{\dagger}\,\text{,}\\
F_{\uparrow\text{L}}^{\dagger}F_{\downarrow\text{R}}  &  =F_{l}F_{s}^{\dagger
}\,\text{,}\\
F_{\downarrow\text{L}}^{\dagger}F_{\uparrow\text{R}}  &  =F_{l}F_{s}%
\,\text{,}\\
F_{\downarrow\text{R}}^{\dagger}F_{\uparrow\text{L}}  &  =F_{s}F_{l}^{\dagger
}\,\text{.} \label{Eq:KleinH}%
\end{align}
Replacing the Klein factors and expanding $H_{\text{K}}^{\perp}$ explicitly
one arrives at%
\end{subequations}
\begin{align}
H_{\text{K}}^{\perp}  &  =\frac{J_{\text{RR}}^{\perp}}{2\pi a}\frac{\tilde
{n}_{c}\tilde{n}_{l}^{+}}{2}\left(  S_{\text{imp}}^{+}F_{s}F_{sl}e^{-i\phi
_{s}}e^{-i\phi_{sl}}\right.  +\nonumber\\
&  \qquad\qquad\qquad+\left.  F_{sl}^{\dagger}F_{s}^{\dagger}e^{i\phi_{s}%
}e^{i\phi_{sl}}S_{\text{imp}}^{-}\right)  +\nonumber\\
&  \quad+\frac{J_{\text{LL}}^{\perp}}{2\pi a}\frac{\tilde{n}_{c}\tilde{n}%
_{l}^{-}}{2}\left(  S_{\text{imp}}^{+}F_{s}F_{sl}^{\dagger}e^{-i\phi_{s}%
}e^{i\phi_{sl}}\right.  +\nonumber\\
&  \qquad\qquad\qquad+\left.  F_{sl}F_{s}^{\dagger}e^{i\phi_{s}}e^{-i\phi
_{sl}}S_{\text{imp}}^{-}\right)  +\nonumber\\
&  \quad+e^{-ieVt}\frac{J_{\text{RL}}^{\perp}}{2\pi a}\frac{\tilde{n}_{c}}%
{2}\left(  \tilde{n}_{sl}^{-}\,S_{\text{imp}}^{+}F_{s}F_{l}^{\dagger}%
e^{-i\phi_{s}}e^{i\phi_{l}}\right.  +\nonumber\\
&  \qquad\qquad\qquad+\left.  \tilde{n}_{sl}^{+}\,F_{s}^{\dagger}%
F_{l}^{\dagger}e^{i\phi_{s}}e^{i\phi_{l}}S_{\text{imp}}^{-}\right)
+\nonumber\\
&  \quad+e^{ieVt}\frac{J_{\text{LR}}^{\perp}}{2\pi a}\frac{\tilde{n}_{c}}%
{2}\left(  \tilde{n}_{sl}^{+}\,S_{\text{imp}}^{+}F_{l}F_{s}e^{-i\phi_{s}%
}e^{-i\phi_{l}}\right.  +\nonumber\\
&  \qquad\qquad\qquad+\left.  \tilde{n}_{sl}^{-}\,F_{l}F_{s}^{\dagger}%
e^{i\phi_{s}}e^{-i\phi_{l}}S_{\text{imp}}^{-}\right)  \,\text{.}%
\end{align}
It is easy to notice that the impurity spin and the lead degrees of freedom
associated with the \textit{spin} sector always appear together in the
combination $S_{\text{imp}}^{+}F_{s}e^{-i\phi_{s}}$ and its Hermitian conjugate.

\subsection{Toulouse Limit and Completion of the Square}

It is natural to describe the strong-coupling limit between the impurity and
the electrons by looking for a transformation that binds those two degrees of
freedom together as a single one. As shown by E\&K, that is achieved by the
following transformation (the boson field is evaluated at the position of the
impurity):%
\begin{equation}
U=e^{i\gamma_{s}\phi_{s}S_{\text{imp}}^{z}}\,\text{.}%
\end{equation}
It follows by simple algebra that $U$ is unitary and commutes with all the
Klein factors and vertex operators (notice that one defines $U$ so that no
\textit{point splitting} is involved when applying it). Using the spin algebra,
$\left[  S_{\text{imp}}^{\pm},S_{\text{imp}}^{z}\right]  =\mp S_{\text{imp}%
}^{\pm}$, and the Baker--Campbell--Hausdorff (BCH) formula, $e^{-B}%
Ae^{B}=A+\left[  A,B\right]  +\frac{1}{2!}\left[  \left[  A,B\right]
,B\right]  +\ldots\,$, we find
\begin{subequations}
\label{0}%
\begin{align}
US_{\text{imp}}^{\pm}U^{\dagger}  &  =S_{\text{imp}}^{\pm}e^{\pm i\gamma
_{s}\phi_{s}}\,\text{,}\\
US_{\text{imp}}^{z}U^{\dagger}  &  =S_{\text{imp}}^{z}\,\text{.}%
\end{align}
Therefore, the perpendicular Kondo term in the Hamiltonian transforms as
$\tilde{H}_{\text{K}}^{\perp}=UH_{\text{K}}^{\perp}U^{\dagger}$, which with
the simplifying choice of $\gamma_{s}=1$ to absorb the spin-sector vertex
into the impurity, and further defining $d^{\dagger}=S_{\text{imp}}^{+}F_{s}$
(so that $d^{\dagger}d=S^{z}+1/2$), takes the form%
\end{subequations}
\begin{align}
\tilde{H}_{\text{K}}^{\perp}  &  =\frac{J_{\text{RR}}^{\perp}}{2\pi a}%
\frac{\tilde{n}_{c}\tilde{n}_{l}^{+}}{2}\left(  d^{\dagger}F_{sl}%
e^{-i\phi_{sl}}+F_{sl}^{\dagger}e^{i\phi_{sl}}d\right)  +\nonumber\\
&  +\frac{J_{\text{LL}}^{\perp}}{2\pi a}\frac{\tilde{n}_{c}\tilde{n}_{l}^{-}%
}{2}\left(  d^{\dagger}F_{sl}^{\dagger}e^{i\phi_{sl}}+F_{sl}e^{-i\phi_{sl}%
}d\right)  +\nonumber\\
&  +e^{-ieVt}\frac{J_{\text{RL}}^{\perp}}{2\pi a}\frac{\tilde{n}_{c}}%
{2}\left(  \tilde{n}_{sl}^{-}\,d^{\dagger}F_{l}^{\dagger}e^{i\phi_{l}}%
-\tilde{n}_{sl}^{+}\,F_{l}^{\dagger}e^{i\phi_{l}}d\right)  -\nonumber\\
&  -e^{ieVt}\frac{J_{\text{LR}}^{\perp}}{2\pi a}\frac{\tilde{n}_{c}}{2}\left(
\tilde{n}_{sl}^{+}\,d^{\dagger}F_{l}e^{-i\phi_{l}}-\tilde{n}_{sl}^{-}%
\,F_{l}e^{-i\phi_{l}}d\right)  \,\text{.}%
\end{align}
Note that no $\tilde{n}_{s}^{\pm}$ factors appeared in these terms; they would
appear in the inter\-lead co\-tunneling terms of $H_{\text{K}}^{z}$, but we
follow S\&H and set the coupling constants of those terms to zero as part of
the definition of the Toulouse limit. So we need to discuss the intra\-lead
part of $H_{\text{K}}^{z}$ and the kinetic terms. Let us examine further the
effects of the E\&K transformation. The transformation of a boson derivative
is given by%
\begin{align}
U\partial\phi_{s}\left(  x\right)  U^{\dagger}  &  =\partial\phi_{s}\left(
x\right)  -i\gamma_{s}S_{\text{imp}}^{z}\left[  \partial\phi_{s}\left(
x\right)  ,\phi_{s}\left(  0\right)  \right]  +\ldots\nonumber\\
&  =\partial\phi_{s}\left(  x\right)  -2\pi\gamma_{s}S_{\text{imp}}^{z}%
\delta\left(  x\right)  \,\text{,}%
\end{align}
where we used $Ae^{B}=e^{B}\left(  A+\left[  A,B\right]  +\ldots\right)  $,
which follows immediately from the BCH formula and the equal-time commutator
$\left[  \phi_{\nu}\left(  x\right)  ,\partial\phi_{\nu^{\prime}}\left(
y\right)  \right]  =2\pi i\delta\left(  x-y\right)  \delta_{\nu\nu^{\prime}}$.
In terms of the corresponding fermions, via debosonization, this shift
translates into a change of boundary conditions and thus $U$ is sometimes
called a \textquotedblleft boundary-condition changing
operator\textquotedblright%
\ \cite{Gogolin,schotte1969,*affleck1994,*ye1997a,*shah2003}. When $\gamma
_{s}=1$ this gives (up to a conventional sign) an Abelian version of the shift
that Affleck and Ludwig use to \textquotedblleft complete the
square\textquotedblright\ and absorb the impurity in a redefinition of the
\textquotedblleft spin density\textquotedblright\ at the infrared fixed point
\cite{affleck1995}. On the one hand, for the spin-sector part of the kinetic
energy, we use this shift and obtain%
\begin{align}
\mathcal{\tilde{H}}_{\nu=s}^{0}  &  =\frac{v_{\text{F}}}{4\pi}\left(
\partial\phi_{s}\left(  x\right)  -2\pi\gamma_{s}S_{\text{imp}}^{z}%
\delta\left(  x\right)  \right)  ^{2}=\\
&  =\frac{v_{\text{F}}}{4\pi}\left[  \partial\phi_{s}\left(  x\right)
\right]  ^{2}-v_{\text{F}}\gamma_{s}S_{\text{imp}}^{z}\partial\phi_{s}\left(
0\right)  +\gamma_{s}^{2}v_{\text{F}}\frac{\pi}{4}\delta\left(  0\right)
\,\text{.}\nonumber
\end{align}
On the other hand, for the parallel Kondo terms we get%
\begin{align}
\tilde{H}_{\text{K}}^{z}  &  =\frac{1}{4\pi}J_{\text{avg}}^{z}S_{\text{imp}%
}^{z}\left[  \partial\phi_{s}\left(  0\right)  -2\pi\gamma_{s}S_{\text{imp}%
}^{z}\delta\left(  0\right)  \right]  +\nonumber\\
&  \quad+\frac{1}{8\pi}\left(  J_{\text{RR}}^{z}-J_{\text{LL}}^{z}\right)
\,S_{\text{imp}}^{z}\partial\phi_{sl}\,\text{,}%
\end{align}
where $J_{\text{avg}}^{z}=\left(  J_{\text{RR}}^{z}+J_{\text{LL}}^{z}\right)
/2$. For the Toulouse limit one considers the symmetric case, $J_{\text{RR}%
}^{z}=J_{\text{LL}}^{z}$, and sets $J_{\text{avg}}^{z}\rightarrow4\pi
v_{\text{F}}\gamma_{s}=4\pi v_{\text{F}}$. Combining these two results and
disregarding constant energy shifts, one has%
\begin{equation}
U\left(  \mathcal{H}_{\nu=s}^{0}+H_{\text{K}}^{z}\right)  U^{\dagger}%
=\frac{v_{\text{F}}}{4\pi}\left[  \partial\phi_{s}\left(  x\right)  \right]
^{2}\,\text{.}%
\end{equation}
In summary, all the parallel Kondo terms were either set to zero or absorbed
into the kinetic term and dropped out from the problem.

Finally, the local-field term is not affected by the transformation procedure
and is written as
\begin{equation}
H_{\text{field}}=-h\,S_{\text{imp}}^{z}=-h\,\left(  d^{\dagger}d-1/2\right)
\,\text{.} \label{Eq:Hh_old}%
\end{equation}

\subsection{Debosonization}

The kinetic terms are easily written back in terms of fermions, becoming
similar to the original kinetic terms of the model. The only nontrivial part
of the Hamiltonian that we need to debosonize and discuss carefully is the
``perpendicular'' Kondo terms. The final result for the lead-symmetric case reads%

\begin{align}
\tilde{H}_{\text{K}}^{\perp}  &  =J_{\text{ra}}\frac{\tilde{n}_{c}\tilde
{n}_{l}^{+}}{2}\left(  d^{\dagger}\psi_{sl}\left(  0\right)  +\psi
_{sl}^{\dagger}\left(  0\right)  d\right)  +\nonumber\\
&  \quad+J_{\text{ra}}\frac{\tilde{n}_{c}\tilde{n}_{l}^{-}}{2}\left(
d^{\dagger}\psi_{sl}^{\dagger}\left(  0\right)  +\psi_{sl}\left(  0\right)
d\right)  -\nonumber\\
&  \quad-J_{\text{er}}\frac{\tilde{n}_{c}\tilde{n}_{sl}^{+}}{2}\left(
d^{\dagger}\psi_{l}\left(  0\right)  +\psi_{l}^{\dagger}\left(  0\right)
d\right)  -\nonumber\\
&  \quad-J_{\text{er}}\frac{\tilde{n}_{c}\tilde{n}_{sl}^{-}}{2}\left(
\psi_{l}^{\dagger}\left(  0\right)  d^{\dagger}+d\psi_{l}\left(  0\right)
\right)\,\text{,}  \label{Eq:Hperp}%
\end{align}
where (cf.~Fig.~\ref{Fig_dot})%
\begin{equation}
J_{\text{ra}}=J_{\ell\ell}^{\perp}/\sqrt{2\pi a}\qquad\text{and}\qquad
J_{\text{er}}=J_{\ell\bar{\ell}}^{\perp}/\sqrt{2\pi a}\,\text{;}%
\end{equation}
and later we will compactly denote%
\begin{equation}
J_{\pm}=\frac{J_{\text{ra}}\tilde{n}_{c}\tilde{n}_{l}^{\pm}}{4v_{\text{F}}%
}\qquad\text{and}\qquad T_{\pm}=\frac{J_{\text{er}}\tilde{n}_{c}\tilde{n}%
_{sl}^{\pm}}{4v_{\text{F}}}\,\text{.}%
\end{equation}
The $\tilde{n}_{\nu}^{\pm}$'s defined in Eq.~(\ref{Eq:Ntwiddles}) are now
viewed as square roots of local fermionic densities at the site of the
impurity. While writing Eq.~(\ref{Eq:Hperp}), we have already gauged away the
time dependence from the coupling constants so that $\mu_{\nu=l}=-(eV)$ and
all other chemical potentials are zero. This is achieved by using the
transformation $\psi_{l}\left(  x,t\right)  =e^{ieV\left(  t-x/v_{\text{F}%
}\right)  }\breve{\psi}_{l}\left(  x,t\right)  $ where $\breve{\psi}_{\nu
}=\frac{1}{\sqrt{2\pi a}}F_{\nu}e^{-i\phi_{\nu}}$ are the debosonized fields,
as was done for the simple junction \cite{shah2016}.

\section{Transport Calculations}

In order to solve for the transport characteristics, we derive an expression
for the current according to $\hat{I}=\partial_{t}\frac{\Delta N}{2}=i\left[
H,\frac{\Delta N}{2}\right]  =i\left[  H_{\text{K}}^{\perp},N_{\nu=l}\right]
$, which gives $I=\left\langle \hat{I}\right\rangle $ as (notice $\left[
\tilde{n}_{l}^{\pm},N_{\nu=l}\right]  =0$)%
\begin{align}
I  &  =-iJ_{\text{er}}\frac{\tilde{n}_{c}}{2}\left[  \tilde{n}_{sl}^{-}\left(
\left\langle d^{\dagger}\psi_{l}^{\dagger}\right\rangle -\left\langle \psi
_{l}d\right\rangle \right)  \right.  -\label{Eq:Current}\\
&  \qquad\qquad\qquad\qquad\qquad-\left.  \tilde{n}_{sl}^{+}\left(
\left\langle \psi_{l}^{\dagger}d\right\rangle -\left\langle d^{\dagger}%
\psi_{l}\right\rangle \right)  \right]  \,\text{.}\nonumber
\end{align}
and the problem reduces to finding those matrix elements.

\subsection{Conventional Approach\label{Sec:Conv}}

In the \textit{conventional} BdB program, the boson exponentials that result
after changing basis are freely recombined and as a result they simply
disappear. This is equivalent, in the expressions above which already
incorporate CBCs that are conventionally not discussed, to replacing
$\tilde{n}_{\nu}^{\pm}\rightarrow1$ everywhere [cf.~Eq.~(\ref{Eq:Ntwiddles})].

To calculate the necessary Green's function elements, using the same 
\textit{local action} scheme and principal-value regularization
(cf.~Ref.~\onlinecite{bolech2004,*bolech2005,*bolech2007,*kakashvili2008a})
as we did for the case of the junction \cite{shah2016}, we adopt the
following Keldysh-Nambu spinor basis (with the frequencies restricted to the
positive semi\-axis only in order to avoid double counting),
\begin{widetext}%
\[
\Psi\left(  \omega\right)  =%
\begin{pmatrix}
\psi_{l}^{-}\left(  \omega\right)  & \psi_{l}^{+}\left(  \omega\right)  &
\psi_{l}^{\dagger-}\left(  \bar{\omega}\right)  & \psi_{l}^{\dagger+}\left(
\bar{\omega}\right)  & d^{-}\left(  \omega\right)  & d^{+}\left(
\omega\right)  & d^{\dagger-}\left(  \bar{\omega}\right)  & d^{\dagger
+}\left(  \bar{\omega}\right)  & \psi_{sl}^{-}\left(  \omega\right)  &
\psi_{sl}^{+}\left(  \omega\right)  & \psi_{sl}^{\dagger-}\left(  \bar{\omega
}\right)  & \psi_{sl}^{\dagger+}\left(  \bar{\omega}\right)
\end{pmatrix}
^{T}\,\text{.}%
\]
Let us define $s_{\nu}\left(  \omega\right)  \equiv\tanh\frac{\omega-\mu_{\nu
}}{2T_{\text{emp}}}$ and $\bar{s}_{\nu}\left(  \omega\right)  \equiv\tanh
\frac{\omega+\mu_{\nu}}{2T_{\text{emp}}}$, where $T_{\text{emp}}$ is the
temperature that is taken to be uniform (and that we will set to zero for the
most part). The local inverse Green's function for the junction,
$G^{-1}\left(  \omega\right)  /2v_{\text{F}}$, is thus given by the following
matrix:%
\[%
\begin{pmatrix}
is_{l} & -is_{l}+i & 0 & 0 & T_{+} & 0 & T_{-} & 0 & 0 & 0 & 0 & 0\\
-is_{l}-i & is_{l} & 0 & 0 & 0 & -T_{+} & 0 & -T_{-} & 0 & 0 & 0 & 0\\
0 & 0 & i\bar{s}_{l} & -i\bar{s}_{l}+i & -T_{-} & 0 & -T_{+} & 0 & 0 & 0 & 0 &
0\\
0 & 0 & -i\bar{s}_{l}-i & i\bar{s}_{l} & 0 & T_{-} & 0 & T_{+} & 0 & 0 & 0 &
0\\
T_{+} & 0 & -T_{-} & 0 & \omega+h & 0 & 0 & 0 & -J_{+} & 0 & -J_{-} & 0\\
0 & -T_{+} & 0 & T_{-} & 0 & -\omega-h & 0 & 0 & 0 & J_{+} & 0 & J_{-}\\
T_{-} & 0 & -T_{+} & 0 & 0 & 0 & \omega-h & 0 & J_{-} & 0 & J_{+} & 0\\
0 & -T_{-} & 0 & T_{+} & 0 & 0 & 0 & -\omega+h & 0 & -J_{-} & 0 & -J_{+}\\
0 & 0 & 0 & 0 & -J_{+} & 0 & J_{-} & 0 & is_{sl} & -is_{sl}+i & 0 & 0\\
0 & 0 & 0 & 0 & 0 & J_{+} & 0 & -J_{-} & -is_{sl}-i & is_{sl} & 0 & 0\\
0 & 0 & 0 & 0 & -J_{-} & 0 & J_{+} & 0 & 0 & 0 & i\bar{s}_{sl} & -i\bar
{s}_{sl}+i\\
0 & 0 & 0 & 0 & 0 & J_{-} & 0 & -J_{+} & 0 & 0 & -i\bar{s}_{sl}-i & i\bar
{s}_{sl}%
\end{pmatrix}
\,\text{;}%
\]

\noindent we also rescaled frequencies and the magnetic field with
$2v_{\text{F}}$ (alternatively, one can think we took $v_{\text{F}}=1/2$).

After finding the Green's functions of interest we replace $T_{\pm}\longmapsto
J_{\text{er}}/2$ and $J_{\pm}\longmapsto\left(  J_{\text{ra}}\pm K\right)
/2$, where in the second one (in order to facilitate the comparison with
previous results from the literature) we reintroduced the possible asymmetry
between right and left leads, which stretching our notation is given by
$K=(J_{\text{ra}}^{\text{L}}-J_{\text{ra}}^{\text{R}})/2$. Combining these
results one gets the following expression for the current:%
\begin{equation}
I=\int_{0}^{+\infty}\frac{J_{\text{er}}^{2}\left[  \left(  J_{\text{er}}%
^{2}+K^{2}\right)  \left(  \omega^{2}+J_{\text{ra}}^{4}\right)  +h^{2}%
J_{\text{ra}}^{2}\right]  \left[  s_{l}\left(  \omega\right)  -\bar{s}%
_{l}\left(  \omega\right)  \right]  }{\omega^{4}+\left[  J_{\text{ra}}%
^{4}+\left(  J_{\text{er}}^{2}+K^{2}\right)  ^{2}-2h^{2}\right]  \omega
^{2}+\left[  J_{\text{ra}}^{2}\left(  J_{\text{er}}^{2}+K^{2}\right)
+h^{2}\right]  ^{2}}\frac{d\omega}{2\pi}\,\text{.} \label{Eq:CurrentSH}%
\end{equation}%
\end{widetext}%

This expression is completely equivalent to the one S\&H reported in their
original work. [See Eqs.~(7) and (8) of Ref.~%
\onlinecite{schiller1995}
and match their notation to ours according to $\Gamma_{a}=\Gamma_{1}+K^{2}$
where $\Gamma_{1}=J_{\text{er}}^{2}$, and $\Gamma_{b}=J_{\text{ra}}^{2}$.] In
particular, it is interesting to consider the case of zero magnetic field. One
finds
\begin{equation}
I=\int_{0}^{+\infty}\frac{J_{\text{er}}^{2}\left(  J_{\text{er}}^{2}%
+K^{2}\right)  }{\omega^{2}+\left(  J_{\text{er}}^{2}+K^{2}\right)  ^{2}%
}\left[  s_{l}\left(  \omega\right)  -\bar{s}_{l}\left(  \omega\right)
\right]  \frac{d\omega}{2\pi}\,\text{,}%
\end{equation}
which is a rather peculiar result, since the dependence on $J_{\text{ra}}$ has
completely dropped out from the problem. This integral is elementary in the
zero-temperature limit,%
\begin{align}
\lim_{T_{\text{emp}}\rightarrow0}I  &  =\frac{1}{\pi}\int_{0}^{V}%
\frac{J_{\text{er}}^{2}\left(  J_{\text{er}}^{2}+K^{2}\right)  }{\omega
^{2}+\left(  J_{\text{er}}^{2}+K^{2}\right)  ^{2}}d\omega\nonumber\\
&  =\frac{1}{\pi}J_{\text{er}}^{2}\arctan\frac{V}{J_{\text{er}}^{2}+K^{2}%
}\,\text{,}%
\end{align}
and it can also be carried out at finite temperature in terms of digamma
functions \cite{schiller1998a}.

Let us focus back on the lead-symmetric case with $K=0$. When there is no
magnetic field, there is no dependence on $J_{\text{ra}}$ in the conventional
result and it applies to the case of $J_{\text{ra}}=0$ in particular. But in
that case the structure of $\tilde{H}_{\text{K}}^{\perp}$ resembles closely
that of the tunneling term in a simple junction \cite{shah2016}, with
$\tilde{n}_{sl}^{\pm}$ playing the role that was played there by
$\tilde{n}_{s}^{\pm}$ and $d^{\dagger}$ replacing $\psi_{sl}^{\dagger}$.
Therefore, in this limit we know (based on our experience with the junction)
how to take the $\tilde{n}$'s into account. And we thus know one should expect
the consistent result to be rather different from the conventional one, except
perhaps for small $J_{\text{er}}$. Indeed, due to the presence of the
$\tilde{n}$ factors in the consistent approach, the structure of the problem
is that of a resonant level attached to a single lead. Because of the
alternation of different $\tilde{n}$'s, there are no Majorana-like operators
that could contribute to the transport and give a nonzero current (cf.~Ref.~%
\onlinecite{roy2012}%
), unlike in the conventional approach. In other words, the expected
consistent-approach result when $J_{\text{ra}}=0$ is simply $dI/dV=0$ for any
value of $J_{\text{er}}$. It follows that we need to treat the $\tilde{n}$'s
consistently for all values of $J_{\text{ra}}$, but that requires additional insights.

\subsection{Consistent Approach}

\begin{figure}[ptb]
\begin{center}
\includegraphics[width=\columnwidth]{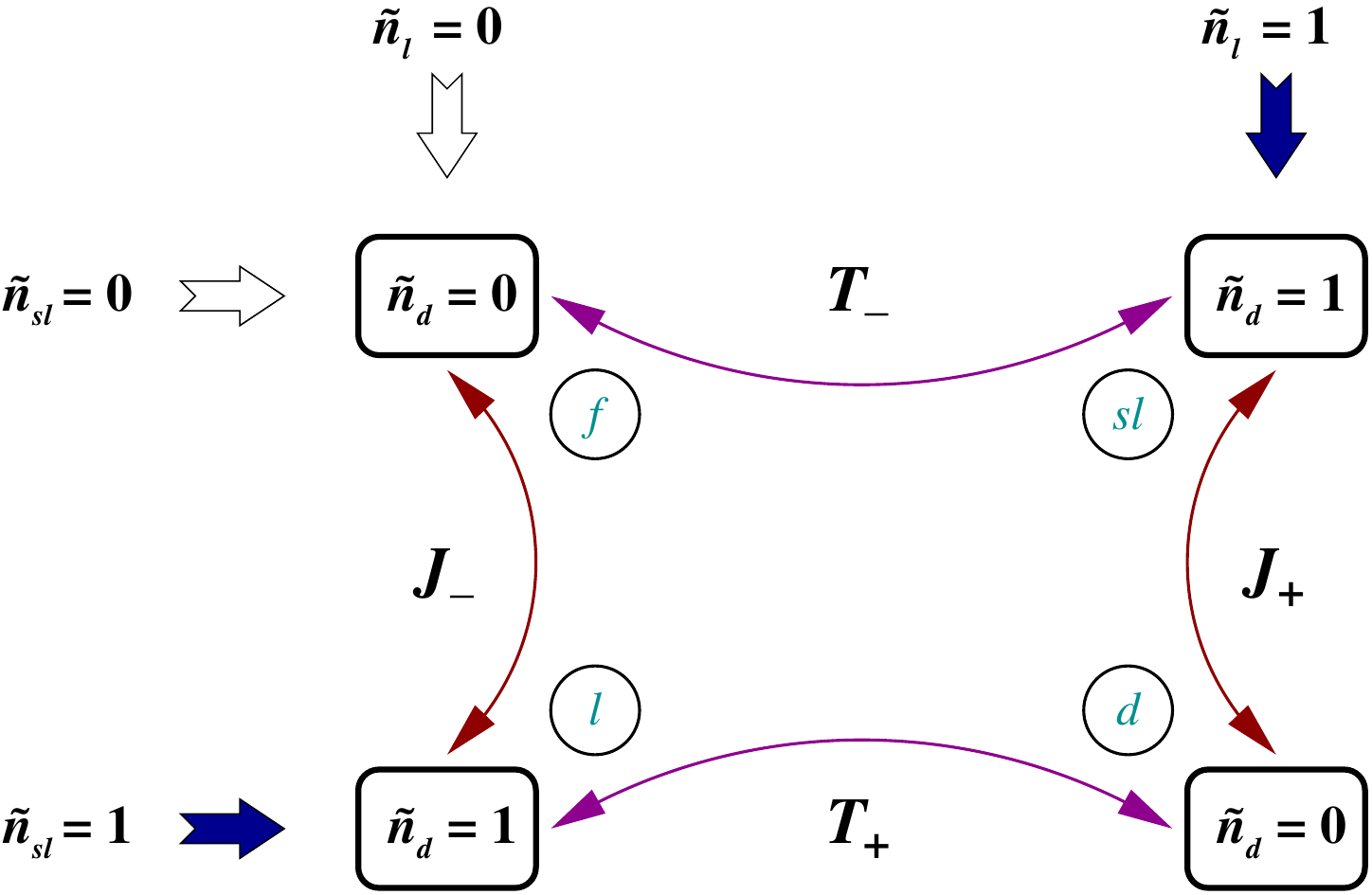}
\end{center}
\caption{Graphical representation of the states that contribute to transport.
Intra\-lead processes are indicated by the vertical arrows ($J_{\pm}$) that
conserve $\tilde{n}_{l}$, while inter\-lead processes are given by the
horizontal arrows ($T_{\pm}$) that conserve $\tilde{n}_{sl}$. The circles
indicate the state labels for the single-particle sector of an equivalent
Gaussian problem that shares the exact same processes (see the discussion in
the text).}%
\label{Fig_states}%
\end{figure}

Let us turn again to the problem of finding the necessary expectation values,
fully dressed by $\tilde{H}_{\text{K}}^{\perp}$ as in Eq.~(\ref{Eq:Hperp}),
needed to compute the current as in Eq.~(\ref{Eq:Current}). The challenge is
to treat the factors $\tilde{n}_{l}^{\pm}$ and $\tilde{n}_{sl}^{\pm}$
consistently. This is a nontrivial problem, but we are nevertheless able to
provide an \textit{ad hoc} solution. In order to achieve that, we start by
studying the local Hilbert space at the impurity site and the structure of the
processes that take place; this is graphically summarized in
Fig.~\ref{Fig_states}.

Let us consider the possible sets of eigen-expectation-values of $\tilde
{n}_{d,l,sl}=0,1$ (cf.~Ref.~%
\onlinecite{shah2016}%
). There are eight combinations in total, but those in which they add up to an
odd number constitute isolated states that are not connected by processes in
$\tilde{H}_{\text{K}}^{\perp}$ and, in particular, do not contribute to
transport. That leaves only four states as depicted in the figure.

From our study of the simple junction \cite{shah2016}, we know that the
physical content of the $\tilde{n}$ factors is actually to avoid contractions
between normal and anomalous terms; in the present case, however, those are
allowed but when and only when $J_{\text{ra}}$ and $J_{\text{er}}$ processes
alternate. By considering processes at different orders of perturbation in
$\tilde{H}_{\text{K}}^{\perp}$, one can conclude that we can achieve the same
set of processes, and also avoid the presence of anomalous terms, by
modifying the \textit{anomalous terms} in Eq.~(\ref{Eq:Hperp}) according to
the following prescription:%
\[
\left\{
\begin{array}
[c]{ccc}%
d^{\dagger} & \longrightarrow & ~~\tilde{f}^{\dagger}\\
\psi_{l}^{\dagger} & \longrightarrow & -\tilde{\psi}_{sl}\\
\psi_{sl}^{\dagger} & \longrightarrow & ~\tilde{\psi}_{l}%
\end{array}
\right.
\]
while not adjusting those same fields (only renaming them by adding twiddles) in the \textit{regular terms}, and
removing all the $\tilde{n}$ factors everywhere. In particular, one can check
processes to fourth order in perturbation, when there are combinations in
which all possible vertexes enter and the state of the system goes around full
circle; cf.~Fig.~\ref{Fig_states}. We verified all these processes are in
one-to-one correspondence in both formulations. This mapping is also argued
for in a different way in the Appendix, by studying some exactly solvable
limits of the two-lead Kondo model.

The mapping has to be applied to a version of $\tilde{H}_{\text{K}}^{\perp}$
previous to that in Eq.~(\ref{Eq:Hperp}), that is already debosonized but in
which the factors of $e^{\pm ieVt}$ still appear explicitly in the inter\-lead
terms and can be removed ulteriorly. We are thus lead to consider%
\begin{align}
\tilde{H}_{\text{K}}^{\perp}  &  =\frac{J_{\text{ra}}^{+}}{2}\left(  \tilde
{d}^{\dagger}\tilde{\psi}_{sl}\left(  0\right)  +\tilde{\psi}_{sl}^{\dagger
}\left(  0\right)  \tilde{d}\right)  +\nonumber\\
&  \quad+\frac{J_{\text{ra}}^{-}}{2}\left(  \tilde{f}^{\dagger}\tilde{\psi
}_{l}\left(  0\right)  +\tilde{\psi}_{l}^{\dagger}\left(  0\right)  \tilde
{f}\right)  -\nonumber\\
&  \quad-\frac{J_{\text{er}}^{+}}{2}\left(  \tilde{d}^{\dagger}\tilde{\psi
}_{l}\left(  0\right)  +\tilde{\psi}_{l}^{\dagger}\left(  0\right)  \tilde
{d}\right)  -\nonumber\\
&  \quad-\frac{J_{\text{er}}^{-}}{2}\left(  \tilde{f}^{\dagger}\tilde{\psi
}_{sl}\left(  0\right)  +\tilde{\psi}_{sl}^{\dagger}\left(  0\right)
\tilde{f}\right)  \,\text{,}%
\end{align}
where the $\pm$ superscripts are used simply to keep track of what was the
corresponding superscript in the now-absent $\tilde{n}_{s}^{\pm}$'s of the
different terms, but in the calculations the two couplings will be taken as
having equal numerical values. The $e^{\pm ieVt}$ factors were removed from
the couplings by using again a time-dependent gauge transformation. The choice
of gauge transformation is this time not unique. We adopted the following
symmetric choice: $\mu_{l}=\mu_{f}=-eV/2$ and $\mu_{sl}=\mu_{d}=eV/2$. Another
choice could have been $\mu_{l}=\mu_{f}=-eV$ and the other two zero; we
checked that these and other choices are all equivalent. We need to stress
that we introduced the $\tilde{\psi}$ notation to emphasize an important
interpretational difference with Eq.~(\ref{Eq:Hperp}): after applying the
prescribed mapping, and even though we kept similar notations for the fields,
there is no connection left to the physical sectors of the theory.

We also want to consider the local magnetic-field term which is naturally
rewritten in a symmetric way (see the Appendix):%
\begin{equation}
\tilde{H}_{\text{field}}=-h\,\left(  \tilde{d}^{\dagger}\tilde{d}+\tilde
{f}^{\dagger}\tilde{f}-1\right)  \,\text{.} \label{Eq:Hh_new}%
\end{equation}
The current is given by%
\begin{align}
I  &  =\left\langle \hat{I}\right\rangle =i\left[  \frac{J_{\text{er}}^{-}}%
{2}\left(  \left\langle \tilde{\psi}_{sl}\tilde{f}^{\dagger}\right\rangle
-\left\langle \tilde{f}\tilde{\psi}_{sl}^{\dagger}\right\rangle \right)
\right.  +\\
&  \qquad\qquad\qquad\qquad\qquad+\left.  \frac{J_{\text{er}}^{+}}{2}\left(
\left\langle \tilde{d}\tilde{\psi}_{l}^{\dagger}\right\rangle -\left\langle
\tilde{\psi}_{l}\tilde{d}^{\dagger}\right\rangle \right)  \right]
\,\text{.}\nonumber
\end{align}
The resulting calculation is straightforward. Using the spinor basis%
\begin{equation}
\Psi\left(  \omega\right)  =%
\begin{pmatrix}
\tilde{\psi}_{sl}^{-} & \tilde{\psi}_{sl}^{+} & \tilde{\psi}_{l}^{-} &
\tilde{\psi}_{l}^{+} & \tilde{d}^{-} & \tilde{d}^{+} & \tilde{f}^{-} &
\tilde{f}^{+}%
\end{pmatrix}
^{T}\,
\end{equation}
and adopting the same notations as for the conventional calculation, the local
inverse Green's function for the junction is given by%
\begin{widetext}%
\[
G^{-1}\left(  \omega\right)  =2v_{\text{F}}%
\begin{pmatrix}
is_{sl} & -is_{sl}+i & 0 & 0 & J_{+} & 0 & -T_{-} & 0\\
-is_{sl}-i & is_{sl} & 0 & 0 & 0 & -J_{+} & 0 & T_{-}\\
0 & 0 & is_{l} & -is_{l}+i & -T_{+} & 0 & J_{-} & 0\\
0 & 0 & -is_{l}-i & is_{l} & 0 & T_{+} & 0 & -J_{-}\\
J_{+} & 0 & -T_{+} & 0 & \omega+h-V & 0 & 0 & 0\\
0 & -J_{+} & 0 & T_{+} & 0 & -\omega-h+V & 0 & 0\\
-T_{-} & 0 & J_{-} & 0 & 0 & 0 & \omega+h+V & 0\\
0 & T_{-} & 0 & -J_{-} & 0 & 0 & 0 & -\omega-h-V
\end{pmatrix}
\,\text{,}%
\]

\noindent where it should be noticed that the voltage now enters explicitly
also in the resonant-level-like diagonal-block action matrix elements
corresponding to $\tilde{d}^{\dagger}$ and $\tilde{f}^{\dagger}$. The voltage
was written here absorbing a factor of $e/2$ only to keep the matrix
expression short; it will be reinserted below. The final expression for the
current is%
\begin{equation}
I=\int_{-\infty}^{+\infty}\frac{d\omega}{2\pi}\frac{2\left(  J_{\text{ss}}%
^{2}-J_{\text{ds}}^{2}\right)  \left(  \omega+h\right)  ^{2}}{\left[  \left(
\omega+h\right)  ^{2}-J_{\text{ds}}^{2}-\left(  eV/2\right)  ^{2}\right]
^{2}+4J_{\text{ss}}^{2}\left(  \omega+h\right)  ^{2}}\left[  s_{l}\left(
\omega\right)  -s_{sl}\left(  \omega\right)  \right]  \,\text{,}
\label{Eq:CurrentInt}%
\end{equation}
where $J_{\text{ss/ds}}=(J_{\text{ra}}^{2}\pm J_{\text{er}}^{2})/4$ are the
sum and the difference of the squares of the couplings, respectively. Notice
that only the squares of the couplings enter into the final expression, which
means that all the $\tilde{n}$'s would have appeared squared as well. As
anticipated, the current vanishes if either $J_{\text{ra}}$ or $J_{\text{er}}$
is zero. The integral can be done in general, but it is simpler and more
illuminating in the zero-temperature limit. Using $d\left(  s_{l}%
-s_{sl}\right)  /dV=\left(  e/2\right)  d\left(  s_{l}-s_{sl}\right)
/d\left(  eV/2\right)  \rightarrow e\left[  \delta\left(  \omega+eV/2\right)
+\delta\left(  \omega-eV/2\right)  \right]  $, we directly write down an
expression for the zero-temperature differential conductance,%
\begin{align}
\frac{dI}{dV}  &  =\sum_{s=\pm1}\frac{2\left(  e/2\pi\right)  \left(
J_{\text{ss}}^{2}-J_{\text{ds}}^{2}\right)  \left(  h+s\,eV/2\right)  ^{2}%
}{\left[  \left(  h+s\,eV/2\right)  ^{2}-J_{\text{ds}}^{2}-\left(
eV/2\right)  ^{2}\right]  ^{2}+4J_{\text{ss}}^{2}\left(  h+s\,eV/2\right)
^{2}}+\nonumber\\
&  +\int_{-\infty}^{+\infty}\frac{d\omega}{2\pi}\frac{4e\left(  eV/2\right)
\left(  J_{\text{ss}}^{2}-J_{\text{ds}}^{2}\right)  \left[  \left(
\omega+h\right)  ^{2}-J_{\text{ds}}^{2}-\left(  eV/2\right)  ^{2}\right]
\left(  \omega+h\right)  ^{2}}{\left\{  \left[  \left(  \omega+h\right)
^{2}-J_{\text{ds}}^{2}-\left(  eV/2\right)  ^{2}\right]  ^{2}+4J_{\text{ss}%
}^{2}\left(  \omega+h\right)  ^{2}\right\}  ^{2}}\left[  s_{l}\left(
\omega\right)  -s_{sl}\left(  \omega\right)  \right]  \,\text{.}%
\end{align}%
\end{widetext}%

\begin{figure}[ptb]
\begin{center}
\includegraphics[width=\columnwidth]{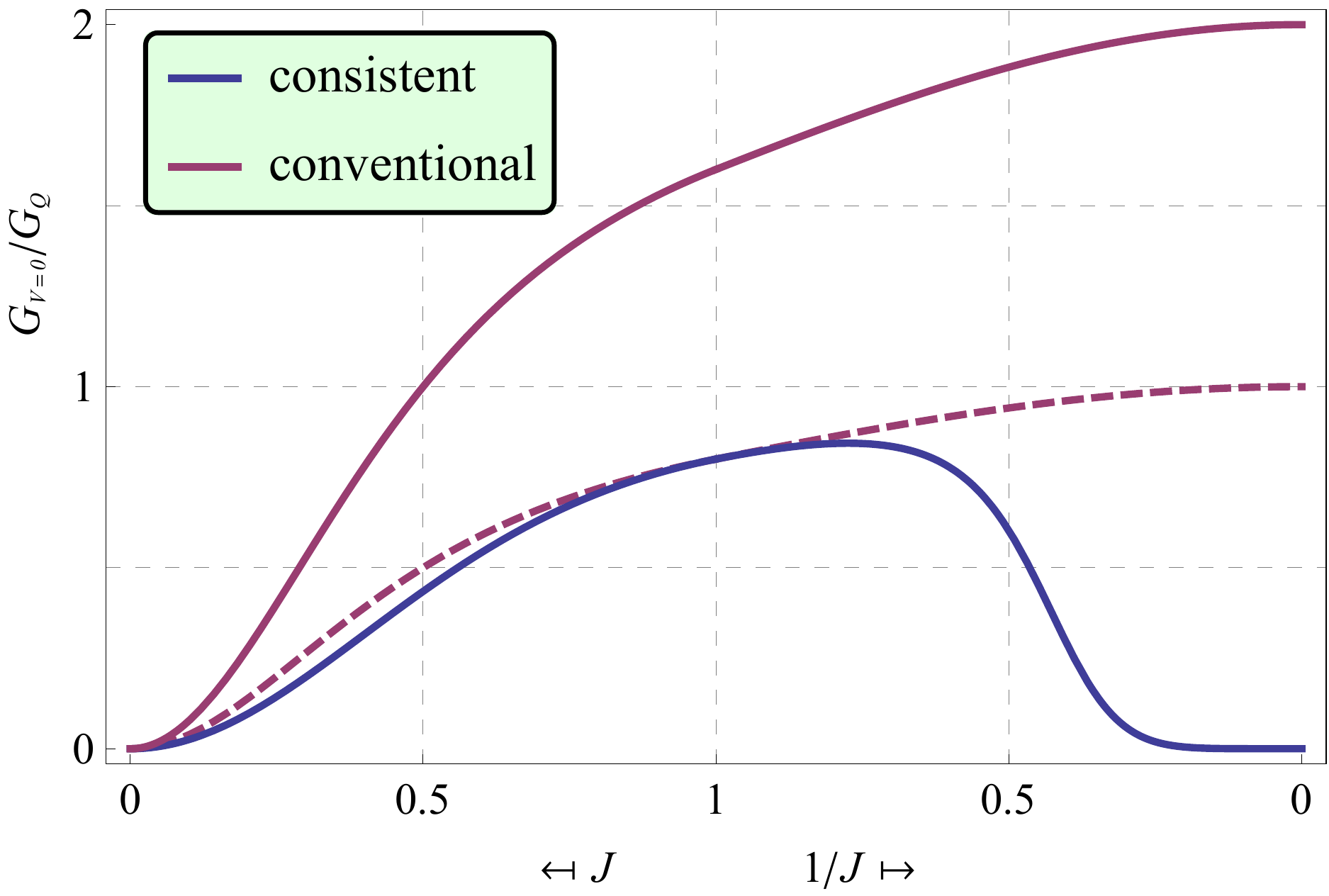}
\end{center}
\caption{Comparison of the differential conductance for the two-lead Kondo
junction, $G=dI/dV$ (in units of the single-channel conductance quantum,
$G_{Q}=e^{2}/h$), calculated according with the conventional procedure or
using our consistent scheme. The plot is at zero applied voltage and finite
magnetic field ($h=v_{\text{F}}$). Notice the convention used for the
horizontal axis in order to cover the full range of $J$ (see also the
corresponding plot for the simple junction \cite{shah2016}). Here, differently
from previous sections, $J$ stands for either $J_{\text{er}}$ or
$J_{\text{ra}}$ while the other ($J_{\text{ra}}$ or $J_{\text{er}}$,
respectively) is set to $1$; the resulting plots are identical for both cases
at zero voltage. The dashed line gives for comparison the conventional result
divided by a factor of $2$.}%
\label{Fig_kondo_dIdV}%
\end{figure}

\subsection{Comparison of Results}

\begin{figure*}[t]
\begin{center}
\includegraphics[width=\columnwidth]{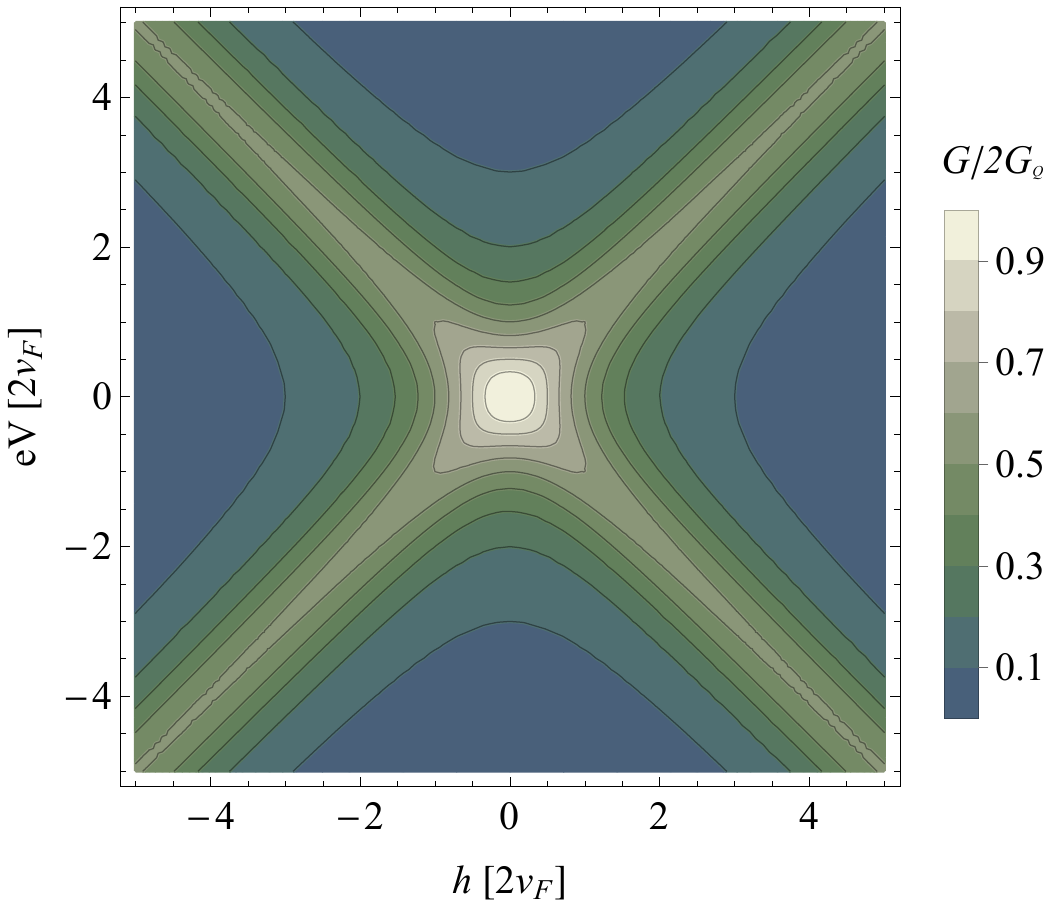}\quad
\includegraphics[width=\columnwidth]{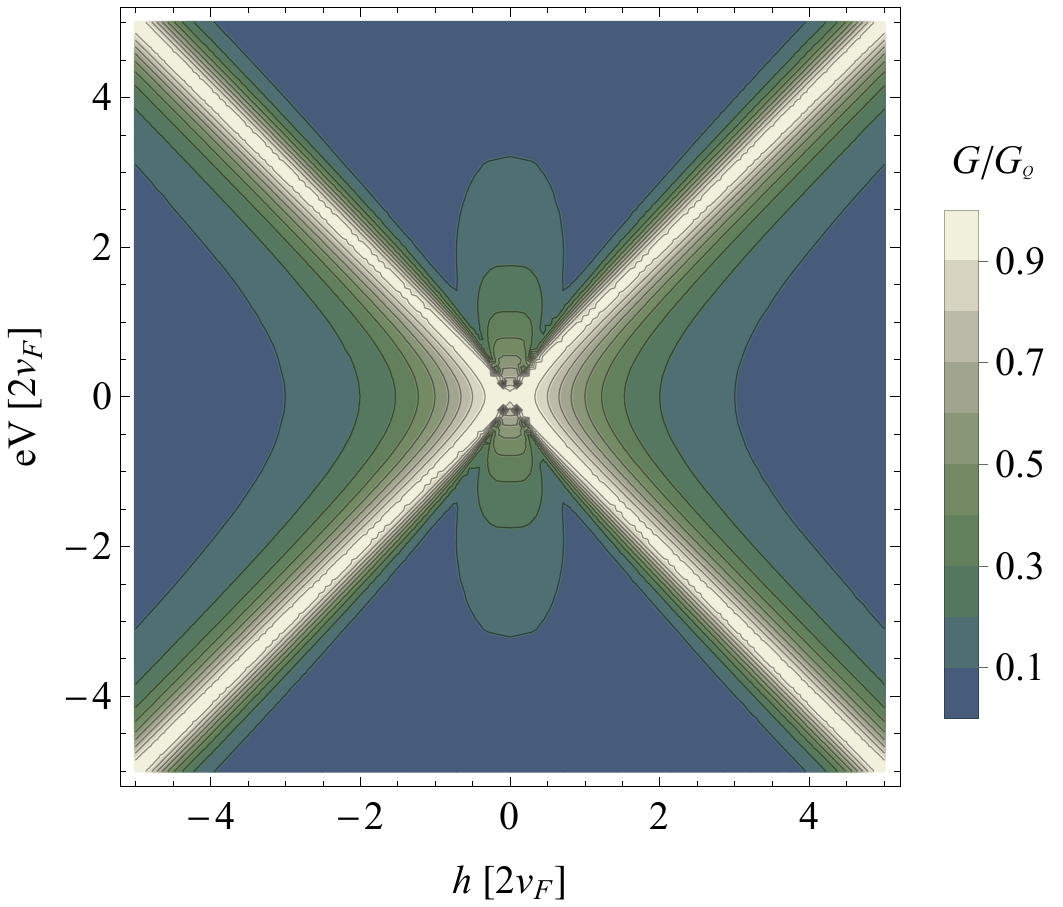}
\end{center}
\caption{Contour color maps of the differential conductance, $G=dI/dV$, in
units of (twice) the single-channel conductance quantum as a function of
applied field and voltage. The two panels compare the conventional calculation
(left panel, in units of $2G_{Q}$) and our consistent calculation (right
panel, in units of $G_{Q}$) for the case of $J_{\text{er}}=1=J_{\text{ra}}$.
While, conventionally, applying a field or a finite voltage have the same
effect, that is not the case in a consistent calculation.}%
\label{Fig_dIdVmaps}%
\end{figure*}

\begin{figure}[t]
\begin{center}
\includegraphics[width=\columnwidth]{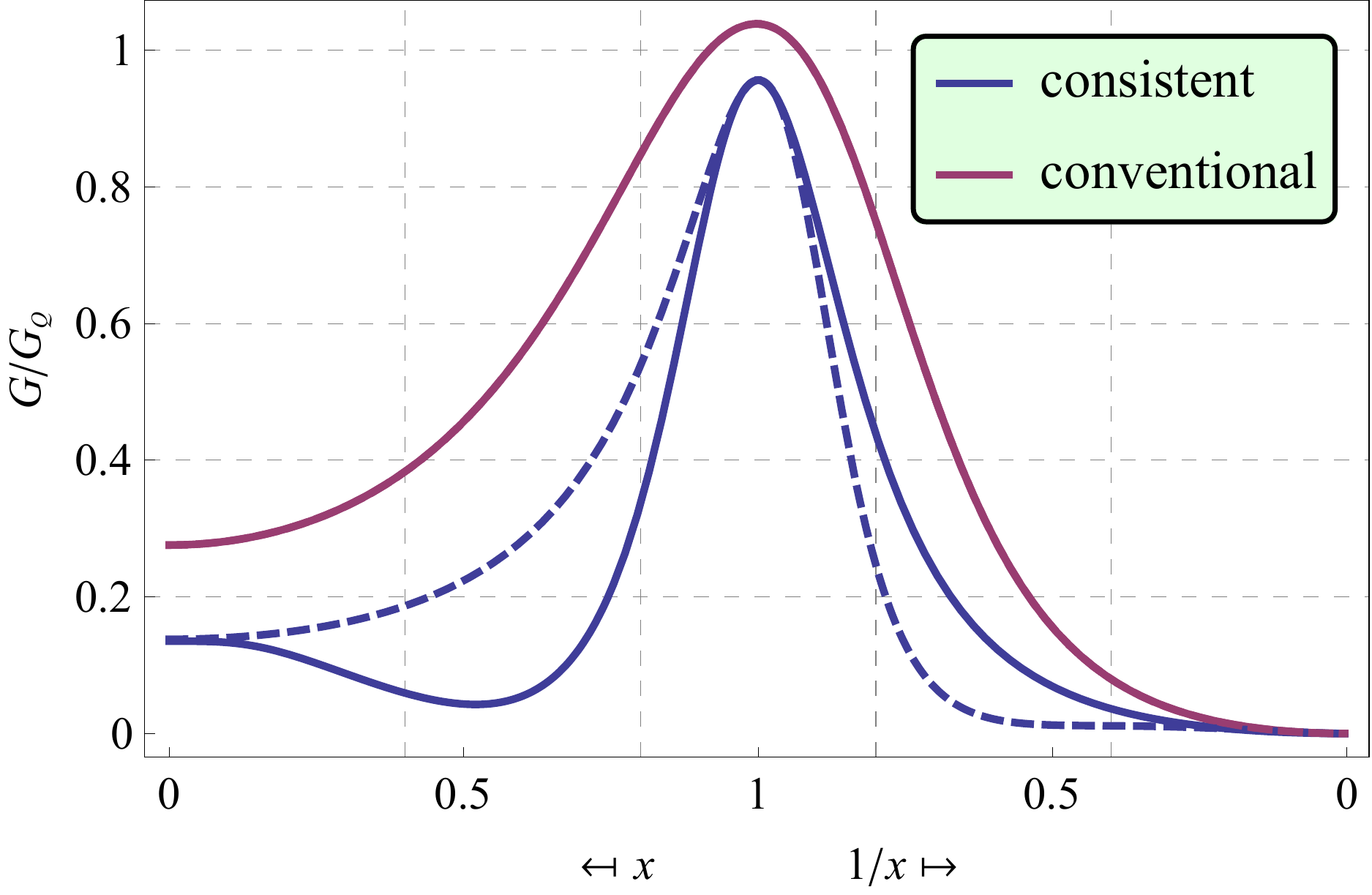}
\end{center}
\caption{Differential conductance for the two-lead Kondo junction, $G=dI/dV$
(in units of the single-channel conductance quantum), calculated according
with the conventional procedure (in red) or using our consistent scheme (in
blue). The plots are either at a finite constant field ($h=5v_{\text{F}}$) and
as a function of applied voltage (dashed lines), or at a finite constant
applied voltage ($eV=5v_{\text{F}}$) and as a function of magnetic field
(solid lines). For the conventional calculation, both plots are identical and
the dashed line is covered by the solid one. Notice the convention used for
the horizontal axis in order to cover the full range of $x$ (defined in terms
of either $x=eV/h$ or $x=h/eV$, with the field that is kept constant being the
one in the denominator).}%
\label{Fig_kondo_dIdVvsVH}%
\end{figure}

\begin{figure}[t]
\begin{center}
\includegraphics[width=\columnwidth]{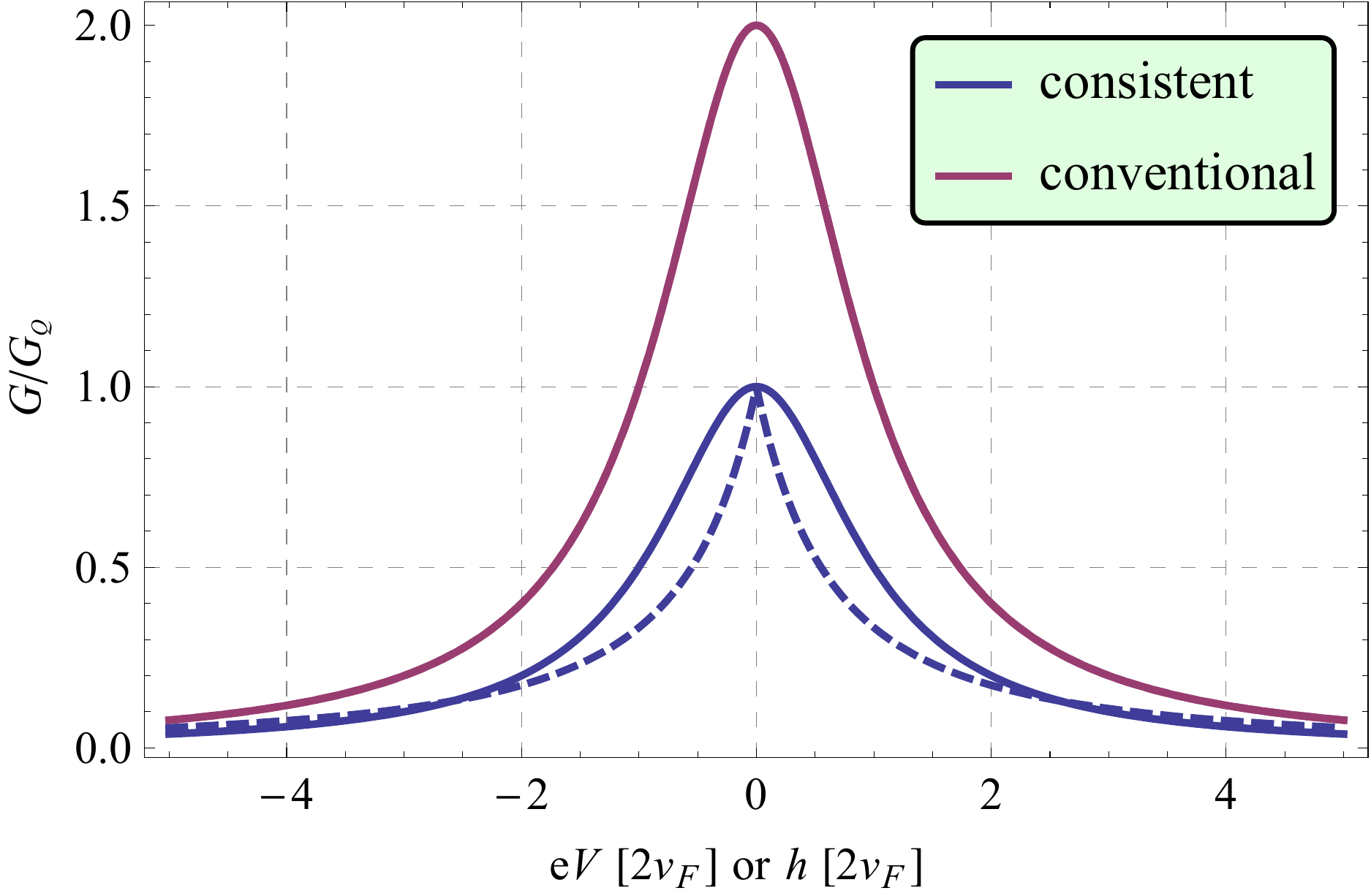}
\end{center}
\caption{Differential conductance for the two-lead Kondo junction, $G=dI/dV$
(in units of the single-channel conductance quantum), calculated according
with the conventional procedure (in red) or using our consistent scheme (in
blue). The plots are either at a zero field and as a function of applied
voltage (dashed lines), or at a zero voltage and as a function of magnetic
field (solid lines). For the conventional calculation, both plots are
identical and the dashed line is covered by the solid one.} 
\label{Fig_kondo_dIdVvsVHzero}%
\end{figure}

In what follows we will illustrate the differences between the results
obtained using the conventional and consistent approaches discussed in the
previous two sections. We will do so mainly by considering the behavior of the
differential conductance, first as a function of $J_{\text{er}}$ and
$J_{\text{ra}}$ for specific values of $h$ and $eV$ as shown in
Fig.~\ref{Fig_kondo_dIdV} and then by looking at its behavior as a function of
$h$ and $eV$ for specific values of $J_{\text{er}}$ and $J_{\text{ra}}$ as
shown in Figs.~\ref{Fig_dIdVmaps} to \ref{Fig_kondo_dIdVvsVHzero}. 
The physical interpretation and plausibility 
of the consistent-approach results will be presented as well.

Figure~\ref{Fig_kondo_dIdV} shows the values of the differential conductance for
voltage $V=0$ and field $h=v_{\text{F}}$ as a function of $J_{\text{er}}$
while fixing $J_{\text{ra}}=1$, or \textrm{vice versa.} We denote
$J_{\text{er}}$ and $J_{\text{ra}}$ by $J\ $in the two cases, respectively.
The two cases produce completely identical plots at zero voltage, as can be
seen from the symmetry of Eqs.~(\ref{Eq:CurrentSH}) and (\ref{Eq:CurrentInt}).
From these expressions, one can also see that for the current and differential
conductance derived using the consistent scheme, the symmetry under exchange
of $J_{\text{er}}$ and $J_{\text{ra}}$ continues to be present also at finite
voltage. On the other hand, the symmetry is absent when one puts $h=0$ for any
value of voltage in the conventional expression and further the transport is
dictated solely by the value of $J_{\text{er}}$. It is to be noted that the
order of limits for $J_{\text{er}}$, $J_{\text{ra}}$, $V$, and $h$ going to
zero is in general important and needs to be treated carefully, as can be seen
by a direct study of the expressions for the current.

From a diagrammatic point of view, one would have expected the conventional
and consistent plots in Fig.~\ref{Fig_kondo_dIdV} to be asymptotically
equivalent for small $J$. An expansion in the couplings of the zero-voltage
$dI/dV$ derived from Eqs.~(\ref{Eq:CurrentSH}) and (\ref{Eq:CurrentInt}) shows
that, unlike the case of a junction \cite{shah2016} or the Ising limit
considered in the Appendix, there is no contribution to the current in the
lowest order. That is the order at which the two calculations would have
matched (as indeed happens for the simple junction and the Ising limit).
The next order is $\mathcal{O}(J_{\text{er}}^{2}J_{\text{ra}}^{2})$, which is
the first non\-vanishing order for the differential conductance and the first
one consistent with the emerging $J_{\text{er}}\leftrightarrow J_{\text{ra}}$
duality of the problem at the Toulouse point that was discussed above. At this
order, the conventional calculation is larger than the consistent one by a
factor of $2$; which is the same factor that one finds for the (same)
next-to-lowest order of the expansion of the consistent calculation as
compared with the exact direct results in the cases of both the junction and
the Ising limit. Interestingly, as can be seen from the comparison of the
$dI/dV$ results for $J_{\text{er}}=J_{\text{ra}}$, the two calculational
schemes yield, at zero voltage, results that are identical except for the
aforementioned factor of $2$. Motivated by these two observations, we also
provided in Fig.~\ref{Fig_kondo_dIdV} the conventional result scaled by a
factor of $1/2$.

Although the tails do not match in Fig.~\ref{Fig_kondo_dIdV}, the two ways of
calculating give the same result as $J$ (\textrm{i.e.}, $J_{\text{er}}$ or
$J_{\text{ra}}$)$\ $goes to zero while the other one (\textrm{i.e.},
$J_{\text{ra}}$ or $J_{\text{er}}$, respectively) is nonzero. However,
similarly to the case of a simple junction \cite{shah2016}, the results are
different for finite $J$, the difference being most marked for large $J$.
While the consistent conductance first increases and then falls down,
eventually going to zero as $J\rightarrow\infty$, the conventional conductance
continues to grow with increasing $J\ $and approaches the value $2e^{2}/h$.

One can gain some additional insights by appealing to physical arguments. The
current is obviously zero if $J_{\text{er}}=0$, but also if $J_{\text{er}%
}\rightarrow\infty$ due to the formation of a resonating-tunneling state that
will block other electrons from approaching the junction (the same as what happens
for a simple junction \cite{shah2016}, since those arguments are not affected
by the fact that tunneling now involves also spin flip). Therefore, a nonzero
current requires the inter\-lead exchange coupling to take some intermediate
value. The physical picture is similar but more involved concerning the
intra\-lead couplings. If $J_{\text{ra}}=0$, for instance, then the parallel
Kondo terms which reach their maximal value at the Toulouse point (measured in
terms of phase shifts) promote the formation of a strongly bound ``static
doublet'' between an electron at each side and the impurity. Such a bound state
``sits'' at the location of the junction and Pauli-blocks the passage of a
current: since electrons tunnel with spin flip, there is always one
bound-state electron already occupying the site with the same spin projection
that the tunneling electron would have either before or after tunneling. If
$J_{\text{ra}}\rightarrow\infty$, in the opposite limit, then a complicated
resonating-exchange state would form instead, but again the current would be
blocked because other electrons would be blocked from approaching the junction
(similarly as for $J_{\text{er}}\rightarrow\infty$). As a result, a
steady-state current requires intermediate values of both inter- and
intra\-lead couplings.

In all the limits that involve strong coupling, the intuitive physical picture
discussed above does not agree with the results of the conventional
calculation, but it does with the consistent calculation in which the
$\tilde{n}$'s are properly taken into account. This is a strong validation for
the need to manipulate and debosonize models consistently after the initial
bosonization. Additionally, it should also be remarked that the consistent
conductance never exceeds the value of one single-channel quantum of
conductance ($G_{Q}\equiv e^{2}/h$). This is also an emergent property at the
non\-equilibrium Toulouse point in addition to the $J_{\text{er}}%
\leftrightarrow J_{\text{ra}}$ transport duality, both of which are indicated
by physical arguments along the lines discussed above. 

It is interesting to take a more systematic look at the variation of the
differential conductance with applied field and voltage. In
Fig.~\ref{Fig_dIdVmaps} we show contour color maps of $dI/dV$ for the
conventional and consistent calculational schemes, while fixing the Kondo
couplings to the symmetric choice $J_{\text{er}}=1=J_{\text{ra}}$. For
additional clarity, in Fig.~\ref{Fig_kondo_dIdVvsVH} we show also two
half-plane cuts of the differential conductance maps: a vertical one at
constant magnetic field, $h/v_{\text{F}}=5$, and a horizontal one at
constant voltage, $eV/v_{\text{F}}=5$. For these cuts, the horizontal axis is
defined in terms of either $x=eV/h$ (dashed lines) or $x=h/eV$ (solid lines),
respectively. As is clearly evident from the figures, the effects of field and
voltage turn out to be exactly equivalent in the conventional
calculation (as seen from the 90$%
{{}^\circ}%
$ rotational symmetry of the first contour plot or from the complete overlap
of the two ``conventional'' traces in the cuts). However, the consistent
calculation yields inequivalent dependencies on field and voltage.

Both ways of calculating show a splitting of the zero-bias differential-conductance 
anomaly due to the finite magnetic field when plotting as a function of voltage, 
but the peaks are sharper and more asymmetric in the consistent calculation. 
The contrast between the two results is even greater at finite applied voltage when
plotting as a function of magnetic field. In the consistent case and for low 
applied voltages, only part of the zero-bias anomaly splits, while a relatively 
broad relic of it remains pinned at zero bias. 
As a result the contour plots are star- or butterfly-shaped, instead of being 
cross-shaped as in the conventional calculation. 
For small $x$ (as compared with $J_{\text{er}}$
and $J_{\text{ra}}$), the differential conductance can additionally be
described as showing a small `deep' developing right after/before the `peak'
when plotting as a function of voltage or field, respectively. In other words,
besides the `ridges' in the contour maps for $\left\vert h\right\vert
=0,\left\vert eV\right\vert $, there are also `furrows' at $\left\vert
2h\right\vert =\left\vert eV\right\vert $ which are absent from the
conventional calculation. This can be seen analytically by looking at the
small-coupling expansion of the differential conductance.

To take a closer look at the central part of the contour maps,
we plot in Fig.~\ref{Fig_kondo_dIdVvsVHzero} horizontal and vertical
cuts going through $h=V=0$ (solid and dashed lines, respectively).
In the conventional calculation, the differential conductance reaches
the maximum value of $2G_{Q}$ at zero field and bias, and decreases
towards zero as either of them increases (once again, the symmetry of
the behavior with $h$ and $V$ make the solid and dashed lines coincide
with each other).
On the other hand, the consistently calculated differential conductance reaches 
a maximum value of $G_{Q}$ at the origin---only half as tall---and it decreases 
towards zero differently with field or with applied voltage, 
with the latter being the slowest decay and the only one of these four curves 
not following a steepest descent.
Notice in addition, from the contour maps, that only along the diagonal ridges
would the differential conductance not go to zero asymptotically.
Let us point out that the way the asymptotic values are approached in all
cases is as a power law instead of the expected logarithmic tail
\cite{kaminski2000,rosch2001}. This is a peculiarity of the Toulouse limit,
in which the voltage or local magnetic field can never be larger than the
parallel Kondo couplings and the bandwidth is infinite. So the standard
argument in which such energy scales stop the renormalization-group
flow of the couplings does not apply. Moreover, already in the conventional
framework and for equilibrium situations, a standardly formulated
renormalization-group scheme \cite{shankar1994} is not compatible with the
nature of the Toulouse-limit fixed-point manifold, which calls for a careful
reformulation \cite{moustakas1996,iucci2008}.

Let us also stress and comment on the differences between the ways in which
applied voltage and temperature enter in both calculations. In the conventional calculation, 
both enter into the expression for the charge current, Eq.~(\ref{Eq:CurrentSH}),
through the thermal factors $s_{l}\left(  \omega\right)$ and $s_{sl}\left(
\omega\right)$ only, but not via the kernel multiplying them (the spectral
function of a Majorana fermion in this case). The latter is independent of
both $eV$ and $T_{\text{emp}}$, which is ascribed to the quadratic nature of
the problem at the solvable point \cite{schiller1998a}. Only as one moves away
from the solvable point, the voltage and temperature will explicitly enter
into the kernel of the integrand \cite{majumdar1998}. This is in
contradistinction to the consistent calculation in which the voltage enters
explicitly into the kernel of Eq.~(\ref{Eq:CurrentInt}), but not the
temperature. Thus, not only voltage and magnetic field, but also voltage and
temperature, interplay differently as compared with the previously accepted,
conventional results for the non\-equilibrium Toulouse point. If one were to imagine
that these kernels capture the Kondo-resonance part of the spectral function
of some parent Anderson-type impurity model, one could see how the resonance
reacts to applied field or voltage. On the one hand, in both calculations the
resonance would shift from zero frequency with magnetic field. On the other hand, 
the resonance would not depend on voltage in the conventional picture, 
but it would split with the applied drain-source bias in the consistent picture. 
The latter would be in agreement with experimental results and the accepted 
phenomenological picture of Kondo transport out of equilibrium 
\cite{defranceschi2002} (and as first predicted in Ref.~\onlinecite{meir1993}).

\section{Conclusion and Final Remarks}

In the companion work of Ref.~%
\onlinecite{shah2016}%
, we presented a \textit{consistent bosonization-debosonization} program in
which we introduced the $\tilde{n}$ factors, defined here in
Eq.~(\ref{Eq:Ntwiddles}), to assist in making the results consistent after
performing transformations in the bosonic language and later debosonizing
models that include terms with single-point non-normal-ordered operators. How
these factors should be treated depends on the physical setup being considered
and needs to be studied on a case by case basis. Sometimes the conventional
way (\textrm{i.e.}, $\tilde{n}\rightarrow1$ for all of them) could be the one
consistent with the problem, but other times the consistent treatment is
different. Moreover, this treatment can be connected to the choice of boundary
conditions that is dictated by the problem (cf.~Ref.~%
\onlinecite{shah2016}%
).

We applied these ideas here to the important case of quantum impurity
problems. In particular, we focused our attention on the two-lead Kondo model
of a junction out of equilibrium. By considering certain regimes of the
problem, we were able to argue that the conventional way of calculating does
not produce consistent results, while a different treatment of the $\tilde{n}$
factors seems to fix those problems (as it did for the case of the simple
junction problem \cite{shah2016}). Moreover, the calculations can then be
carried out in the full regime of parameters of the system. This way, the key
insights of the work by S\&H (and also by E\&K) can be retained and the
calculations fixed to produce consistent results (our method of solution was
\textit{ad hoc} and there is no exact solution to refer to as in the case of
the junction, but we do know it interpolates between consistent limits). We
thus were able to make a number of predictions for the transport
characteristics of the two-lead Kondo model that can, in principle, be looked
for in experiments.

Certainly, more work should be done along this line following the developments
of the literature of the past two decades, and we are already exploring some
directions. To name but a few: (i) the consistent solution can be explored
further, including additional aspects of transport (such as the noise spectrum
and possibly thermal transport), alternatives such as charge sensing
\cite{berman1999,*bolech2005b}, and also the thermodynamics; (ii) one can study
ac-drive effects that are realizable in experiments
\cite{schiller1996,kogan2004a,*kogan2004b}; (iii) a study of multi\-terminal
models would provide new insights, cf.~Refs.~%
\onlinecite{Gogolin,fabrizio1994,*tsvelik1995,*shah2006}%
; (iv) perturbation around the solvable Toulouse point needs to be considered
anew, cf.~Refs.~%
\onlinecite{clarke1993,*sengupta1994,majumdar1998}%
; (v) connections to other approaches such as boundary CFT can bring in synergy
\cite{affleck1990,*affleck1991,johannesson2003,*johannesson2005}, showing for
example how to possibly extend those methods to non\-equilibrium transport
problems; (vi) \textrm{ditto} for approaches that exploit the connections to
integrability \cite{konik2002,*mehta2006} or to renormalization ideas
\cite{anders2005a,*anders2008,*schmitt2010}; (vii) one could combine our approach with a
finite-size bosonization analysis to further bridge with CFT and numerical
renormalization ideas \cite{vondelft1998,*zarand2000}. The list can go on
(even though we restricted it to quantum impurity problems only). Given the
continued challenge posed by the need to better understand strongly correlated
quantum systems out of equilibrium, the consistent Toulouse-point solution
will play an important role as a reference case for a class of problems in
which a lot is still not well understood and there is a lack of exact results
to guide the theoretical developments.

\begin{acknowledgments}
We acknowledge the hospitality of the Kavli Institute for Theoretical Physics
and the Aspen Center for Physics where part of this research was done. C.J.B.
thanks A.~Schiller for several past discussions of his work on
quantum-impurity physics. N.S. acknowledges the hospitality of the University of
Cologne while she studied Refs.~%
\onlinecite{vondelft1998a,schiller1998a,zarand2000}
and thanks A.~Rosch for providing her an opportunity to present a series of
pedagogical blackboard talks on that topic.
\end{acknowledgments}

\appendix*

\section{Some Exactly Solvable Limits}

In the following, we consider two different limits that can be rigorously
treated via exact direct calculations. They further motivate the prescription
mapping discussed in the main text to solve for the transport within the consistent approach.

\subsection{$\hat{x}$-axis Ising Limit}

We call the Ising limit of the anisotropic Kondo model that in which all the
spin-spin exchange interactions take place along a single axis. We will choose
it to be the  (as is often done for the transverse-field Ising model).
The virtue of this limit is that it is exactly solvable: the electrons still
interact with the impurity but the latter does not have dynamics
(\textrm{i.e.}, no spin flips take place along the $\hat{x}$ axis). As a result, all
one has to do is to solve for each possible impurity orientation and average
over the results (this is reminiscent of the treatment of the boundary
sine-Gordon model that one obtains at the solvable point of the problem of a
classical impurity in a Tomonaga-Luttinger liquid
\cite{Gogolin,kane1992,*kane1992a}; cf.~Ref.~%
\onlinecite{guinea1985,*guinea1985a}%
). When the impurity is frozen, the only remaining degrees of freedom are the
electronic ones and the problem becomes Gaussian and thus exactly solvable in
a direct way. We shall keep the $\hat{z}$ axis as the quantization axis for the
electrons, as this will show some structure that will help us understand how
to deal with the full two-lead Kondo model in the language of Abelian bosonization.

In the zero-field case, and setting in Eq.~(\ref{Eq:2LK}) all the non-$x$
couplings to zero, $J_{\ell\ell^{\prime}}^{y},J_{\ell\ell^{\prime}}^{z}=0$, we
are left with (the time dependence is implicit)
\begin{subequations}
\label{Eq:2LxIK}%
\begin{align}
H  &  =\sum_{\sigma,\ell}\left(  \int\mathcal{H}_{\ell}^{0}\,dx+H_{\text{K}%
}^{x}\right)  \,\text{,}\\
\mathcal{H}_{\ell}^{0}  &  =\psi_{\sigma\ell}^{\dagger}\left(  x\right)
\left(  -iv_{\text{F}}\partial_{x}\right)  \psi_{\sigma\ell}\left(  x\right)
\,\text{,}\\
H_{\text{K}}^{x}  &  =J_{\ell\ell^{\prime}}^{x}\,S_{\text{imp}}^{x}\left(
\frac{1}{2}\psi_{\bar{\sigma}\ell}^{\dagger}\left(  0\right)  \psi_{\sigma
\ell^{\prime}}\left(  0\right)  \right)  \,\text{,}%
\end{align}
where we have made use of the relation $2S_{\text{imp}}^{x}S_{\text{elec}}%
^{x}=S_{\text{imp}}^{x}\left(  S_{\text{elec}}^{+}+S_{\text{elec}}^{-}\right)
$. Hermiticity requires $J_{\text{RL}}^{x}=J_{\text{LR}}^{x}=J_{\text{er}}%
^{x}$. It is a simple exercise to solve the model directly and find a closed
expression for the differential conductance (for brevity, we do not quote that
result here).

We are interested in bosonizing, changing boson basis, and debosonizing in the
\textit{same way} as we did for the full model (gauging the voltage in and out
from the couplings also in the same way). Of course, no unitary transformation
is required this time since $S_{\text{imp}}^{x}$ commutes with the Hamiltonian
and thus plays a simple spectator role. The Kondo part of the Hamiltonian is
finally rewritten as (all fields are at time $t$ and $x=0$)%
\end{subequations}
\begin{align}
H_{\text{K}}^{x}  &  =J_{\text{RR}}^{x}\frac{\tilde{n}_{c}\tilde{n}_{l}^{+}%
}{2}S_{\text{imp}}^{x}\left(  \psi_{sl}^{\dagger}\psi_{s}^{\dagger}+\psi
_{s}\psi_{sl}\right)  -\nonumber\\
&  \quad-J_{\text{LL}}^{x}\frac{\tilde{n}_{c}\tilde{n}_{l}^{-}}{2}%
S_{\text{imp}}^{x}\left(  \psi_{s}^{\dagger}\psi_{sl}+\psi_{sl}^{\dagger}%
\psi_{s}\right)  +\nonumber\\
&  \quad+J_{\text{er}}^{x}\frac{\tilde{n}_{c}\tilde{n}_{sl}^{+}}%
{2}S_{\text{imp}}^{x}\left(  \psi_{s}^{\dagger}\psi_{l}^{\dagger}+\psi_{l}%
\psi_{s}\right)  -\nonumber\\
&  \quad-J_{\text{er}}^{x}\frac{\tilde{n}_{c}\tilde{n}_{sl}^{-}}%
{2}S_{\text{imp}}^{x}\left(  \psi_{l}^{\dagger}\psi_{s}+\psi_{s}^{\dagger}%
\psi_{l}\right)  \,\text{.} \label{Eq:HxKold}%
\end{align}

One could now set all $\tilde{n}\rightarrow1$ and proceed conventionally to
calculate the differential conductance. The result is that the expression
differs from the direct calculation in a way that parallels what we discussed
for a simple junction \cite{shah2016}. In particular, a small-coupling
expansion shows that the results match to lowest order, $\mathcal{O}%
[(J_{\text{er}}^{x})^{2}]$, but the conventional result is twice bigger than
the direct one at next-leading order, as was the case for the simple junction.
Therefore, one needs to treat the $\tilde{n}$'s more carefully.

The practical problem that arises is that the factors $\tilde{n}_{l}^{\pm}$
and $\tilde{n}_{sl}^{\pm}$ introduce complicated dynamics into the Hamiltonian
because there are linear terms in $\psi_{l}^{\left[  \dagger\right]  }$ and
$\psi_{sl}^{\left[  \dagger\right]  }$ present as well. From our study of the
simple junction \cite{shah2016}, we know that the physical content of these
factors is actually to avoid contractions between regular and anomalous terms
of the same type (\textrm{i.e.}, inter- or intra\-lead). Here the situation is
more complicated because regular and anomalous terms of a given type can be
combined provided there is an intervening term of the other type, and
\textrm{vice versa}. One can check this conclusion order by order via a
matching of perturbative expansions as we did for the case of the simple
junction \cite{shah2016}.

But since the present problem is directly solvable in terms of the original
fermions, we know there exists a way of organizing the perturbation theory as
if the model was purely Gaussian. We shall thus focus on the structure of the
terms while comparing new and original fermions. We do that by looking at the
four Klein factor relations [see Eqs.~(\ref{Eq:KleinA})-(\ref{Eq:KleinH})] that
involve $F_{s}^{\dagger}$: (i) for the regular terms, $F_{sl}F_{s}^{\dagger
}=F_{\uparrow\text{L}}^{\dagger}F_{\downarrow\text{L}}$ and $F_{l}%
F_{s}^{\dagger}=F_{\uparrow\text{L}}^{\dagger}F_{\downarrow\text{R}}$; and
(ii) for the anomalous terms, $F_{sl}^{\dagger}F_{s}^{\dagger}=F_{\uparrow
\text{R}}^{\dagger}F_{\downarrow\text{R}}$ and $F_{l}^{\dagger}F_{s}^{\dagger
}=-F_{\uparrow\text{R}}^{\dagger}F_{\downarrow\text{L}}$. The other four are
just the Hermitian conjugates of these.

A first observation is that in terms of the original fermions there are no
anomalous terms present in the model; see Eq.~(\ref{Eq:2LxIK}). This prompts
us to attempt a mapping in which we do not change the regular terms but
`regularize' the anomalous ones. Notice that while $F_{s}^{\dagger}$ appears
in all four terms, no single original Klein factor appears four times; while
$F_{\uparrow\text{L}}^{\dagger}$ repeats in the regular terms, $F_{\uparrow
\text{R}}^{\dagger}$ does in the anomalous ones.\ A second observation is that
the pair of spin-down original Klein factors repeat in regular and anomalous
terms, but they exchange roles as to which one goes in the intra\-lead process
and which in the inter\-lead process in each case.

These two observations indicate that we can achieve the same perturbative
processes (and, as a bonus, avoid the presence of anomalous terms) by
modifying \textit{only} the anomalous terms according to the following
prescription:%
\begin{equation}
\left\{
\begin{array}
[c]{ccc}%
\psi_{s}^{\dagger} & \longrightarrow & ~~\tilde{\psi}_{z}^{\dagger}\\
\psi_{l}^{\dagger} & \longrightarrow & -\tilde{\psi}_{sl}\\
\psi_{sl}^{\dagger} & \longrightarrow & ~\tilde{\psi}_{l}%
\end{array}
\right.
\end{equation}
and removing all the $\tilde{n}$ factors. To keep the same phase conventions
as we used for the matching of Klein factor bilinears, we also need to
introduce a minus sign in all the terms (this is not essential as the current
is not sensitive to it). Explicitly, one has%
\begin{align}
H_{\text{K}}^{x}  &  =\frac{1}{2}J_{\text{R}}^{x}S_{\text{imp}}^{x}\left(
\tilde{\psi}_{l}^{\dagger}\tilde{\psi}_{z}+\tilde{\psi}_{z}^{\dagger}%
\tilde{\psi}_{l}\right)  +\nonumber\\
&  \quad+\underset{~}{\frac{1}{2}}J_{\text{L}}^{x}S_{\text{imp}}^{x}\left(
\tilde{\psi}_{s}^{\dagger}\tilde{\psi}_{sl}+\tilde{\psi}_{sl}^{\dagger}%
\tilde{\psi}_{s}\right)  +\nonumber\\
&  \quad+\overset{~}{\frac{1}{2}}J_{\text{er}}^{x}S_{\text{imp}}^{x}\left(
\tilde{\psi}_{sl}^{\dagger}\tilde{\psi}_{z}+\tilde{\psi}_{z}^{\dagger}%
\tilde{\psi}_{sl}\right)  +\nonumber\\
&  \quad+\frac{1}{2}J_{\text{er}}^{x}S_{\text{imp}}^{x}\left(  \tilde{\psi
}_{l}^{\dagger}\tilde{\psi}_{s}+\tilde{\psi}_{s}^{\dagger}\tilde{\psi}%
_{l}\right)  \label{Eq:HxKnew}%
\end{align}
and, trivially since the $\tilde{\psi}$'s can be put in one-to-one
correspondence with the original fermions in Eq.~(\ref{Eq:2LxIK}), all the
results will be the same as in the direct solution. We stress that, even
though we kept the notation with $s$, $l$, and $sl$, these are different
fermions and there is no direct connection left to the physical sectors of the
theory (we introduced the $\tilde{\psi}$ notation to emphasize this point).

\subsection{Flat-band Limit}

If one introduces a lattice discretization, the flat-band limit is the limit
of zero hopping between sites. Then all sites in each band are independent
fermionic degrees of freedom, except for the sites at $x_{0}$, the location of
the impurity, which are connected by the tunneling terms in the Hamiltonian.
We could consider this limit for the Toulouse-point Hamiltonian with its Kondo
part given by Eq.~(\ref{Eq:Hperp}) or for the BdB-mapped $x$-axis-Ising
Hamiltonian corresponding to Eq.~(\ref{Eq:HxKold}). Both yield equivalent
models in the flat-band limit; for brevity, we frame our presentation around
the second case. Taking the band energies to be zero, we introduce the
notation $\psi_{\nu}^{\dagger}\left(  x_{0}\right)  \rightarrow c_{\nu
}^{\dagger}$ for $\nu=s,l,sl$ (or $d^{\dagger}\rightarrow c_{s}^{\dagger}$ in
the first case) to emphasize the lattice nature of the problem and the sector
of the Hamiltonian containing $x=x_{0}$ turns into the following three-site
model (taking the expectation value of the ``spectator'' impurity spin and
absorbing it in redefined coupling constants),%
\begin{align}
H_{\text{3s}}  &  =J_{\text{ra}}^{\text{a}}n_{l}^{+}\left(  c_{sl}^{\dagger
}c_{s}^{\dagger}+c_{s}c_{sl}\right)  -J_{\text{ra}}^{\text{r}}n_{l}^{-}\left(
c_{s}^{\dagger}c_{sl}+c_{sl}^{\dagger}c_{s}\right)  +\nonumber\\
&  \quad+J_{\text{er}}^{\text{a}}n_{sl}^{+}\left(  c_{s}^{\dagger}%
c_{l}^{\dagger}+c_{l}c_{s}\right)  -J_{\text{er}}^{\text{r}}n_{sl}^{-}\left(
c_{l}^{\dagger}c_{s}+c_{s}^{\dagger}c_{l}\right)  \,\text{.}%
\end{align}
Since now we are dealing with lattice fermions, we can directly identify
$\tilde{n}_{\nu}\rightarrow n_{\nu}$.

This model has an eight-state Hilbert space split into two
particle-number-parity sectors. The states with even number of particles are
all disconnected degenerate states with zero energy. The states with odd
particle number hybridize and can be diagonalized (using exact
diagonalization) into four eigenstates with energies $\pm\sqrt{-b/2\pm
\sqrt{b^{2}/4-c}}$ where $b=-\left[  \left(  J_{\text{ra}}^{\text{a}}\right)
^{2}+\left(  J_{\text{ra}}^{\text{r}}\right)  ^{2}+\left(  J_{\text{er}%
}^{\text{a}}\right)  ^{2}+\left(  J_{\text{er}}^{\text{r}}\right)
^{2}\right]  $ and $c=\left(  J_{\text{ra}}^{\text{a}}J_{\text{ra}}^{\text{r}%
}-J_{\text{er}}^{\text{a}}J_{\text{er}}^{\text{r}}\right)  ^{2}$.

Applying the prescribed changes for the anomalous terms, this Hamiltonian
turns into the equivalent of Eq.~(\ref{Eq:HxKnew}), which is now a four-site
model with no anomalous terms,%
\begin{align}
H_{\text{4s}}  &  =-J_{\text{ra}}^{\text{a}}\left(  \tilde{c}_{z}^{\dagger
}\tilde{c}_{l}+\tilde{c}_{l}^{\dagger}\tilde{c}_{z}\right)  -J_{\text{ra}%
}^{\text{r}}\left(  \tilde{c}_{s}^{\dagger}\tilde{c}_{sl}+\tilde{c}%
_{sl}^{\dagger}\tilde{c}_{s}\right)  -\nonumber\\
&  \quad-J_{\text{er}}^{\text{a}}\left(  \tilde{c}_{z}^{\dagger}\tilde{c}%
_{sl}+\tilde{c}_{sl}^{\dagger}\tilde{c}_{z}\right)  -J_{\text{er}}^{\text{r}%
}\left(  \tilde{c}_{l}^{\dagger}\tilde{c}_{s}+\tilde{c}_{s}^{\dagger}\tilde
{c}_{l}\right)  \,\text{.}%
\end{align}
This model conserves particle number and, in the single-particle sector, its
spectrum coincides with the odd sector of $H_{\text{3s}}$
(cf.~Fig.~\ref{Fig_states}). Moreover, the same conclusion remains true when
we introduce a local magnetic field as in Eqs.~(\ref{Eq:Hh_old}) or
(\ref{Eq:Hh_new}), appropriately rewritten. This serves as a check of how to
correctly normalize the local-field term after the mapping. In other words,
the Hamiltonian matrices in those two sectors are unitary equivalent,%
\begin{equation}
P_{\text{odd}}H_{\text{3s}}P_{\text{odd}}\underset{U}{\equiv}P_{\text{1p}%
}H_{\text{4s}}P_{\text{1p}}\,\text{.}%
\end{equation}

Due to the absence of interactions in $H_{\text{4s}}$, its many-particle
physics can be calculated in terms of a Green's function for a single fermion
injected into an empty band \cite{Doniach}, and thus the single-particle
sector is the crucial one to determine the full dynamics. Notice, in addition,
that the way this kind of Green's functions at the bare level enter into the
transport calculations of the main text is via their inverses. As such, the
propagator of any hybridized local or flat-band degree of freedom will not
require independent regularization and will inherit its causal properties
(\textrm{i.e.}, its Keldysh structure
\cite{bolech2004,*bolech2005,*bolech2007,*kakashvili2008a}) from other
extended-band degrees of freedom; and thus the difference between ``empty-band''
and ``degenerate-gas'' Green's functions does not enter the calculations. Thus,
for those physical properties whose calculation requires only the two-point
Green's functions and, by extension, only the single-particle sector, the
Hamiltonian mapped via the prescription should work fine. And in the
particular example above we were able to show explicitly that the
single-particle sector of the mapped model reproduces the nontrivial part of
the original spectrum.

\bigskip

In summary, the mapping we prescribed for the anomalous terms works in both
cases, namely the linear- and the flat-band limits. In the case of the
two-lead-Kondo-model Toulouse point, we encountered two linear-band and one
local degrees of freedom, which is a combination of these two cases we just
discussed. One can resort to neither bosonization nor exact diagonalization to
prove the mapping rigorously, but it is nevertheless justified on physical
grounds as a combination of these two limits and as argued also in terms of
processes in the main text.

\bibliography{astrings,books2015,kondo2015,hubbard2015,tunneling2015}

\end{document}